\documentclass[12pt]{article}

\newcommand{\ra}{\rightarrow}
\newcommand{\nn}{\nonumber}

\newcommand{\al}{\alpha}
\newcommand{\be}{\begin{equation}}
\newcommand{\ee}{\end{equation}}

\newcommand{\ov}{\overline}

\newcommand{\diag}{{\rm diag}}

\newsavebox{\ns}
\newsavebox{\dbrane}
\newsavebox{\dbshort}

\def \Tr{{\rm Tr}}
\def\*{\star}
\def\({\left(}		
\def\){\right)}		
\def\[{\left[}		
\def\]{\right]}

\def\CM{{\cal M}}	\def\CN{{\cal N}}	
\def\CP{{\cal P}}

\title{\vspace{-1in}\parbox{\linewidth}{\small\hfill\shortstack{IASSNS-HEP-98/77\\
ITEP-TH-40/98\\hep-th/9808175}}
\vspace{0.6in}\\
New ${\cal N}=2$ superconformal field theories from M/F theory orbifolds}
\author{Sergei Gukov\thanks{On leave from Princeton University, Princeton,
NJ 08544, USA}\\
{\sl\small ITEP, Bol.Cheremushkinskaya, 25, Moscow 117259 Russia}\\
{\sl\small Landau Institute for Theoretical Physics, Kosygina 2, Moscow 117940 Russia}\and
Anton Kapustin\\
{\sl\small School of Natural Sciences, Institute for Advanced Study}\\
{\sl\small Olden Lane, Princeton, NJ 08540}}

\begin{document}
\begin{titlepage}
\renewcommand{\thepage}{ }
\renewcommand{\today}{ }
\thispagestyle{empty}

\maketitle

\begin{abstract}
We consider M-theory on $({\bf T}^2\times {\bf R}^2)/{\bf Z}_n$
with M5 branes wrapped on ${\bf R}^2$. One can probe this background with M5
branes wrapped on ${\bf T}^2$. The theories on the probes provide many new
examples of ${\cal N}=2$ field theories without Lagrangian description. 
All these theories have Coulomb branches, and we find the corresponding Seiberg-Witten
curves. The exact solution is encoded in a Hitchin system on an orbifolded torus
with punctures. The theories we consider also arise from D3 probes in F-theory
on K3$\times$K3 orbifolds. Interestingly, the relevant F-theory background has frozen
${\bf Z}_n$ singularities which are analogous to frozen ${\bf Z}_2$ singularities in
Type I string theory. We use the F-theory description to find
supergravity duals of the probe SCFT's in the large N limit and compute the spectrum
of relevant and marginal operators. We also explain how the decoupling of $U(1)$ factors
is manifested in the supergravity description.

\end{abstract}
\end{titlepage}

\section{Introduction}
In the last few years many examples of nontrivial superconformal field
theories (SCFT) in $d=4$ have been found. In particular, there exists a uniform way to
construct ${\cal N}=2$ SCFT's by looking for singular points on the Coulomb branch of 
${\cal N}=2$ field theories. Although we do not yet have a complete understanding of these 
SCFT's, one can learn a lot about them from the low-energy effective action away from the 
singular points (i.e. from the Seiberg-Witten solution)~\cite{SWAP}. In addition, when an
SCFT has a large N limit, it can often be described by IIB supergravity on
$AdS^5\times X$ for a suitable $X$~\cite{maldacena, AdS}.

In this paper we are going to describe and study a class of ${\cal N}=2$
SCFT's which arise from consideration of M5 probes on M-theory orbifolds.
The M-theory background we consider is $({\bf T}^2\times {\bf R}^2)/{\bf Z}_n$, $n
=2,3,4,6$, where the ${\bf Z}_n$ action preserves sixteen supersymmetries. 
For $n=3,4,6$ the complex structure of ${\bf T}^2$ must be restricted appropriately. 
We will think of $({\bf T}^2\times {\bf R}^2)/{\bf Z}_n$ as an elliptic
fibration over ${\bf R}^2/{\bf Z}_n$. The M-theory background will also include
$k$ M5 branes wrapping the noncompact base of this fibration. 
The resulting configuration has eight unbroken supersymmetries. The probe M5's are 
wrapping the elliptic fiber at $N$ points of the base. We will find it
convenient to work on the $n$-fold cover of the base ${\bf R}^2$. Then, counting images,
we have $nN$ probe M5's wrapped at $nN$ points of ${\bf  R}^2$. The probes do not
break supersymmetry further.

For a ${\bf Z}_2$ orbifold ($n=2$) the $\CN=2$ theory on the probes has a Lagrangian
description.
To see this, one takes the $\tau$-parameter of the fiber to infinity. In this limit
the M-theory background described above reduces to a
IIA configuration containing two $O6^-$ planes, with two D6 branes on top of
each of them, and $k$ NS5 branes. (The reason one gets two D6 branes on top of each 
of the $O6^-$ planes is that in such a configuration the Ramond-Ramond charge is cancelled
locally. Hence this IIA configuration lifts to a locally flat M-theory geometry,
namely an orbifold geometry.) The probe M5 branes reduce to D4 branes suspended between
NS5 branes. Such ``elliptic'' brane configurations have been previously considered
in~\cite{Angel}. Elliptic brane configurations do not allow for the bending of NS5 branes
and therefore always produce theories with vanishing beta-functions~\cite{Witten}.
In the present case the low-energy theory on D4 branes has a product gauge group with both
symplectic and unitary factors~\cite{Angel}. The gauge groups and matter content are 
the same as in $d=6$ theories arising on the worldvolume of Type I D5 branes near a 
${\bf Z}_{2k}$ singularity~\cite{DM}.

When $n=3,4,6$ the $\tau$-parameter of the fiber is a root of unity, so no IIA
description exists. As a consequence, the
probe theories do not have a Lagrangian description. However, by taking the
volume of the fiber to zero, one may pass to an F-theory description.
The dual F-theory background is $({\bf T}^2\times {\bf R}^2)/{\bf Z}_n\times
{\bf C^2}/{\bf Z}_{nk}$, and the probe M5 branes become probe D3 branes.
Note that the F-theory background can be regarded as an orbifold limit of K3$\times$K3,
therefore it has eight unbroken supersymmetries. Introduction of D3 probes does not
break supersymmetry further. This F-theory background can be also thought
of as a collection of several mutually nonlocal 7-branes wrapped on 
${\bf C^2}/{\bf Z}_{nk}$. The IIB coupling is constant
over the base and equal to a root of unity. The case $k=0$ is especially simple
since the 7-branes become flat. In this case the theory on the
7-branes has a gauge group $E_6,E_7,$ or $E_8$ depending on the value of $n$~\cite{DasMu}. This
implies that the theory on the probe D3 branes has an exceptional global symmetry  
and does not admit a Lagrangian description. The latter presumably remains true for
$k>0$. Thus we obtain three infinite families of new ${\cal N}=2$ SCFT's without 
Lagrangian description labeled by $n=3,4,6$.

An important subtlety is that the F-theory orbifold corresponding to
M-theory on $({\bf T}^2\times{\bf R}^2)/{\bf Z}_n$ with $k$ M5 branes is not a geometric
orbifold. By this we mean that some of the blow-up modes of the ${\bf C}^2/{\bf Z}_{nk}$
orbifold are absent. Concretely, we will show  that the
${\bf Z}_{nk}$ singularity on which the 7-branes are wrapped
can only be resolved to a product of $k$ ${\bf Z}_n$ singularities. These ``frozen''
${\bf Z}_n$ singularities cannot be further resolved.
For $n=2$ the ``frozen'' singularity is T-dual to the ``frozen'' singularity
of Type I on ${\bf C}^2/{\bf Z}_2$~\cite{P}. For $n>2$ we obtain a
nonperturbative generalization of~\cite{P}. This is discussed in more detail
in section 3. 

To study new SCFT's we will make use both of M-theory and F-theory descriptions. 
The M-theory setup is convenient for finding Seiberg-Witten solutions. In the next
section we show that for all $n$ and $k$ the solution is encoded in a Hitchin system on 
an orbifolded torus with punctures.  From this we derive the Seiberg-Witten curve 
(the spectral cover of the Hitchin system) and the S-duality group of the probe 
theories. As a matter of fact, the Hitchin system solves not just the theory in $d=4$, 
but its compactification on a circle of arbitrary radius~\cite{K}. The F-theory description, 
discussed in detail in section 3, is helpful for finding supergravity duals of the SCFT's in 
the limit of large number of probes. In section 4 we use these supergravity duals to compute 
the spectrum of relevant and marginal operators in the SCFT's. Most of the spectrum can be checked independently
using our knowledge of the Seiberg-Witten solution,
along the lines of~\cite{SWAP}. We find complete agreement between the two approaches.
We also discuss the decoupling of $U(1)$ factors in the boundary gauge theory
(in the cases where the boundary SCFT is equivalent to a gauge theory). We show
that from the supergravity point of view this effect arises from a certain
subtlety in the Kaluza-Klein reduction of a 2-form on $AdS^5\times S^1$.
The details of this argument are explained in the Appendix. 
Our conclusions are summarized in section 5. 

\section{Seiberg-Witten curves from M theory orbifolds}
Consider M-theory on $({\bf T}^2\times {\bf R}^2)/{\bf Z}_n$
with $k$ M5 branes wrapped on ${\bf R}^2/{\bf Z}_n$
and $N$ M5 branes wrapped on the elliptic fiber ${\bf T}^2$ at $N$ points of 
${\bf R}^2/{\bf Z}_n$ (or $nN$ points of ${\bf R}^2$, if we work on the cover). 
$n$ is one of the integers $2,3,4,6$. 
For $n=3,6$ the $\tau$-parameter of ${\bf T}^2$ has
to be $\exp(i\pi/3)$, for $n=4$ $\tau=i$, and for $n=2$ $\tau$ is unrestricted.
If we denote the complex affine coordinates on ${\bf T}^2$ and ${\bf R}^2$ by
$z$ and $v$, then the ${\bf Z}_n$ action is
\be
z\ra \omega z,\qquad v\ra \omega^{-1} v,
\ee
where $\omega=\exp(2\pi i/n)$. This orbifolding preserves sixteen
supersymmetries. The M5 branes wrapped on ${\bf R}^2$ break half of the
remaining supersymmetries. The probe M5 branes wrapping ${\bf T}^2$ 
do not break any further supersymmetries. To distinguish the two sets of M5 branes
we will call the branes wrapping ${\bf R}^2$ the M5$'$ branes.

\subsection{${\bf Z}_2$ orbifold}
We start with brane configurations which have a IIA limit. The corresponding probe
theories are described by $\CN=2$ gauge theories in the infrared. The IIA limit exists
when the $\tau$-parameter of the torus can be taken to infinity, so we have to set $n=2$.
Let the two coordinates on ${\bf T}^2$ be $x^6,x^{10}$, and the coordinates
on ${\bf R}^2$ be $x^4,x^5$. The probe M5 branes 
wrap $x^6,x^{10}$, while M5$'$ branes wrap $x^4,x^5$. Upon reduction to IIA the orbifold
background in M-theory reduces to $({\bf R}^2\times {\bf S}^1)/(\Omega (-1)^{F_L}{\cal R}_{456})$,
where $S^1$ is parametrized by $x^6$, $\Omega$ is worldsheet parity, and ${\cal R}_{456}$ 
is the reflection of $x^4,x^5,x^6$.
The action of ${\cal R}_{456}$ has two fixed planes: $x^4=x^5=x^6=0$ and
$x^4=x^5=0,x^6=\pi R_6$. These fixed planes are the $O6^-$ planes of IIA.
There are also two D6 branes on top of each of the orientifold planes.\footnote{To see
this consider a D2 brane probing this background. The theory on the D2 brane located
near one of the $O6^-$ planes is an ${\cal N}=4$ $d=3$ $SU(2)$ theory with two
fundamentals. Its moduli space is an orbifold $({\bf R}^3\times {\bf S}^1)/{\bf
Z}_2$~\cite{SW3}.
Therefore the corresponding M-theory background is also a ${\bf Z}_2$ orbifold.}
The M5 branes reduce to $2N$ D4 branes stretched along $x^6$. The M5$'$
branes reduce to $k$ NS5 branes extended in $x^4,x^5$ and localized in $x^6$.
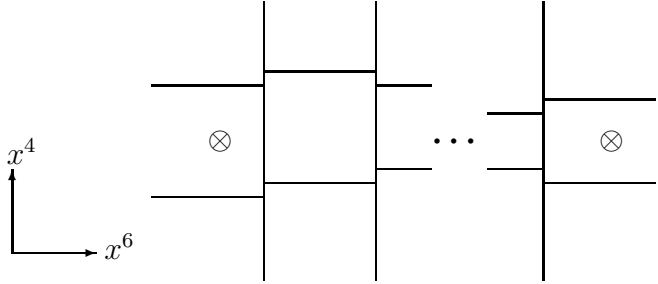
\begin{figure}

\setlength{\unitlength}{0.9em}
\begin{center}
\begin{picture}(22,8)

\savebox{\ns}(0.2,8)[l]{\line(0,1){10}}

\savebox{\dbrane}(4,0){\line(1,0){4}}
\savebox{\dbshort}(2,0){\line(1,0){2}}

\put(6,3.73){$\otimes$}\put(20,3.73){$\otimes$}
\put(8,0){\usebox{\ns}}\put(12,0){\usebox{\ns}}\put(18,0){\usebox{\ns}}
\put(4,6){\usebox{\dbrane}}\put(4,2){\usebox{\dbrane}}
\put(8,6.5){\usebox{\dbrane}}\put(8,2.5){\usebox{\dbrane}}
\put(12,6){\usebox{\dbshort}}\put(12,3){\usebox{\dbshort}}
\multiput(14.2,4)(0.6,0){3}{\circle*{.2}}
\put(16,5){\usebox{\dbshort}}\put(16,3){\usebox{\dbshort}}
\put(18,5.5){\usebox{\dbrane}}\put(18,2.5){\usebox{\dbrane}}

\put(-1,0){\vector(1,0){3}}\put(-1,0){\vector(0,1){3}}
\put(2.3,-0.2){$x^6$}\put(-1.2,3.2){$x^4$}

\end{picture}\end{center}
\caption{Elliptic IIA brane configuration. Vertical and horizontal
lines are NS5 branes and D4 branes, respectively. $\otimes$ denotes an
$O6^-$ plane with two D6 branes on top. Only half of the circle parametrized by
$x^6$ is shown.}
\end{figure}
The resulting brane
configuration is shown in Figure 1. It appeared previously in~\cite{Angel}. It
preserves eight supersymmetries, therefore the theory on D4 branes is an ${\cal N}=2$
$d=4$ theory at low energies. Well-known arguments~\cite{Witten} show that its gauge 
group ${\cal G}$ is $Sp(N)\times SU(2N)^{k-1}\times Sp(N)$. Naively, the gauge group is
$Sp(N)\times U(2N)^{k-1}\times Sp(N)$, but the $U(1)$ factors decouple~\cite{Witten}.
To state the matter content, we
arrange the simple factors of ${\cal G}$ along a line. There is a bifundamental
hypermultiplet for every pair of neighboring group factors. The 
two $Sp(N)$ factors in addition have two fundamentals each. The resulting gauge theory has 
zero beta-functions and is an example of a finite ${\cal N}=2$ theory. We will show later
that this theory has an interesting S-duality group. Theories in $d=6$ with precisely this 
gauge group and matter content have previously appeared in the study of D5 probes near 
a ${\bf Z}_{2k}$ singularity in Type I~\cite{DM,IB}. This is not a coincidence: T-duality 
along $x^6,x^4,x^5$ converts our setup to that studied in~\cite{DM,IB}.

\begin{figure}

\setlength{\unitlength}{0.9em}
\begin{center}
\begin{picture}(22,8)

\savebox{\ns}(0.2,8)[l]{\line(0,1){10}}

\savebox{\dbrane}(4,0){\line(1,0){4}}
\savebox{\dbshort}(2,0){\line(1,0){2}}

\put(3.57,3.73){$\otimes$}\put(20,3.73){$\otimes$}
\put(4,0){\usebox{\ns}}
\put(8,0){\usebox{\ns}}\put(12,0){\usebox{\ns}}\put(18,0){\usebox{\ns}}
\put(4,7){\usebox{\dbrane}}\put(4,2){\usebox{\dbrane}}
\put(2,6){\usebox{\dbshort}}\put(2,1){\usebox{\dbshort}}
\put(8,6.5){\usebox{\dbrane}}\put(8,2.5){\usebox{\dbrane}}
\put(12,6){\usebox{\dbshort}}\put(12,3){\usebox{\dbshort}}
\multiput(14.2,4)(0.6,0){3}{\circle*{.2}}
\put(16,5){\usebox{\dbshort}}\put(16,3){\usebox{\dbshort}}
\put(18,5.5){\usebox{\dbrane}}\put(18,2.5){\usebox{\dbrane}}

\put(-3,0){\vector(1,0){3}}\put(-3,0){\vector(0,1){3}}
\put(0.3,-0.2){$x^6$}\put(-3.2,3.2){$x^4$}

\end{picture}\end{center}
\caption{Elliptic IIA brane configuration with an unpaired NS5 brane on top
of one of the orientifold planes. The notation is the same as in Figure 1.}
\end{figure}
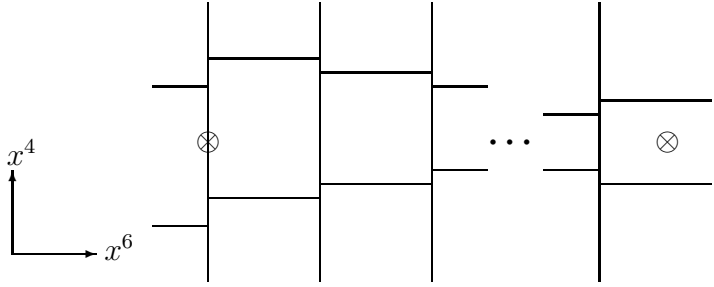

It is also possible to put unpaired NS5 branes on top of
orientifold planes. The resulting brane configurations are shown in Figures 2 and 3.
In Figure 2, we have $2k$ pairs of NS5 branes and one unpaired NS5 brane.
The latter
can move in the $x^7,x^8,x^9$ but not in $x^6,x^{10}$. The theory on D4 branes
has gauge group $SU(2N)^k\times Sp(N)$. There are bifundamental hypers for each
pair of neighboring simple factors. The first $SU(2N)$ factor has in addition a
hypermultiplet in the two-index antisymmetric representation and two fundamentals,
and the $Sp(N)$ factor has two fundamentals. In Figure 3, we have $2k$
paired NS5 branes
and two unpaired ones at each of the orientifold planes. The gauge group is
$SU(2N)^{k+1}$, with $k$ bifundamentals. The first and last factor 
also have an antisymmetric tensor and two fundamentals. It is easy to check
that all these gauge theories are finite.

\begin{figure}

\setlength{\unitlength}{0.9em}
\begin{center}
\begin{picture}(22,10)

\savebox{\ns}(0.2,8)[l]{\line(0,1){10}}

\savebox{\dbrane}(4,0){\line(1,0){4}}
\savebox{\dbshort}(2,0){\line(1,0){2}}

\put(3.57,3.73){$\otimes$}\put(21.57,3.73){$\otimes$}
\put(4,0){\usebox{\ns}}\put(22,0){\usebox{\ns}}
\put(8,0){\usebox{\ns}}\put(12,0){\usebox{\ns}}\put(18,0){\usebox{\ns}}
\put(4,7){\usebox{\dbrane}}\put(4,2){\usebox{\dbrane}}
\put(2,6){\usebox{\dbshort}}\put(2,1){\usebox{\dbshort}}
\put(8,6.5){\usebox{\dbrane}}\put(8,2.5){\usebox{\dbrane}}
\put(12,6){\usebox{\dbshort}}\put(12,3){\usebox{\dbshort}}
\multiput(14.2,4)(0.6,0){3}{\circle*{.2}}
\put(16,5){\usebox{\dbshort}}\put(16,3){\usebox{\dbshort}}
\put(18,5.5){\usebox{\dbrane}}\put(18,1.5){\usebox{\dbrane}}
\put(22,2.5){\usebox{\dbshort}}\put(22,6.5){\usebox{\dbshort}}

\put(-3,0){\vector(1,0){3}}\put(-3,0){\vector(0,1){3}}
\put(0.3,-0.2){$x^6$}\put(-3.2,3.2){$x^4$}

\end{picture}\end{center}
\caption{Elliptic IIA brane configuration with two unpaired NS5 branes.
The notation is the same as in Figure 1.}
\end{figure}
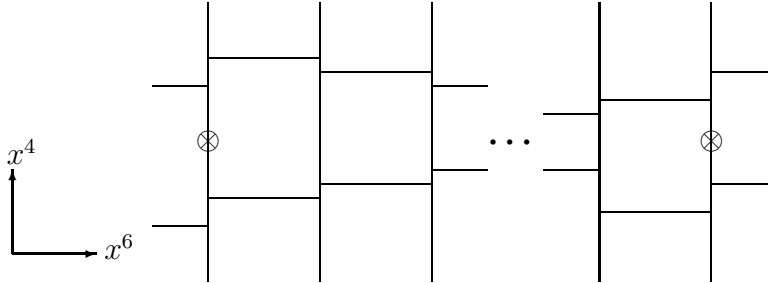

\begin{sloppypar}
Flavor symmetries acting on the fundamental hypermultiplets are 
$(Spin(4)\times Spin(4))/{\bf Z}_2$, $(U(2)\times Spin(4))/{\bf Z}_2$, 
and $(U(2)\times U(2))/{\bf Z}_2$ for configurations in Figures 1,2, and 3, respectively. 
The ${\bf Z}_2$ quotient arises because part of the naive flavor symmetry is
gauged. There is also a global $U(1)$ for each bifundamental and
antisymmetric tensor. The case $k=0$ is special, since the gauge group
becomes simple. Indeed, for $k=0$ both Figures 1 and 2
correspond to an $Sp(N)$ gauge theory with an antisymmetric tensor and four
fundamentals,
while Figure 3 yields an $SU(2N)$ gauge theory with two antisymmetric
tensors and four fundamentals. Correspondingly, the 
flavor symmetry acting on the fundamentals
is enhanced to $Spin(8)/{\bf Z}_2$ in the former case and to $U(4)/{\bf Z}_2$ 
in the latter.
\end{sloppypar}

The way we constructed our brane configuration does not allow to have nonzero masses for
fundamental hypermultiplets. These masses become nonzero if the D6 branes do not coincide 
with $O6^-$ planes. However in such a configuration the Ramond-Ramond charge is no longer 
cancelled locally. The corresponding M-theory background 
is curved and is not described by an orbifold. 
On the other hand, the masses for the bifundamentals and the antisymmetric tensors
are allowed by the orbifold geometry. They correspond to the differences 
in the center-of-mass positions of the neighboring stacks of D4 branes~\cite{Witten}. 
Because of this, the sum of all the masses must be zero: $\sum_\al m_\al=0$. 
As in~\cite{Witten}, this condition can be relaxed by
introducing a shift in $x^4,x^5$ as one goes around $x^6$. The corresponding mass 
parameter is referred to as the global mass~\cite{Witten,Angel}.

Let us consider the effect of T-duality along $x^6$. The $O6^-$ planes and
four D6 branes become an $O7^-$ plane and four D7 branes, with a $Spin(8)/{\bf Z}_2$ 
gauge theory on their worldvolume. This worldvolume is not flat: four of its coordinates 
are wrapped on an ALF space with an orbifold singularity resulting from T-dualizing 
NS5 branes. For configurations in Figures 1,2, and 3 this singularity is ${\bf Z}_{2k}$,
${\bf Z}_{2k+1}$, and ${\bf Z}_{2k+2}$, respectively. There is a subtlety
here: as noticed by Polchinski~\cite{P}, in the presence of an orientifold projection there 
is a possibility of having frozen ${\bf Z}_2$ singularities, with no blow-up modes. This
phenomenon is analogous to discrete torsion in oriented string theory~\cite{VVW}. 
As a matter of fact, many blow-up modes {\it are} frozen in our situation. Namely, the
configuration T-dual to Figure 1 has only enough blow-up modes for a resolution 
${\bf Z}_{2k}\ra ({\bf Z}_2)^k$, the one T-dual to Figure 2 can be blown-up as 
${\bf Z}_{2k+1}\ra ({\bf Z}_2)^k$,
and the one T-dual to Figure 3 as ${\bf Z}_{2k+2}\ra ({\bf Z}_2)^k$. To see this, note that
the blow-up modes correspond to the motion of NS5 branes in $x^7,x^8,x^9$.
However, an NS5 brane and its mirror image always move together, so the corresponding 
${\bf Z}_2$ singularity cannot be resolved. The unpaired NS5 branes can move independently.
This simple observation leads immediately to the ``freezing'' pattern described above.

A ``frozen'' ${\bf Z}_2$ singularity can be ``thawed'' by bringing an NS5
brane and its image to an orientifold plane and then separating them in the $x^7,x^8,x^9$.
After this procedure we loose the zero mode corresponding to the motion along $x^6,x^{10}$,
but gain a zero mode corresponding to the separation of the NS5 branes in $x^7,x^8,x^9$.
At the transition point
the NS5 brane and its image are coincident, so there are tensionless strings. 
After T-duality in $x^6$ this phase transition can be described as follows. We are dealing
with D7 branes wrapped on a ``frozen'' ${\bf C}^2/{\bf Z}_2$ orbifold. The low-energy
$d=4$ theory is lacking a blow-up mode (a hypermultiplet), but has an extra vector
multiplet. We will call this branch of the moduli space the Coulomb branch. 
At a special point on the Coulomb branch we may pass to a Higgs 
branch, where the vector multiplet becomes massive and the blow-up mode reappears. 
The $d=6$ version of this transition was considered in~\cite{IB,HZ}. In particular,
\cite{HZ} gives a description of the $d=6$ transition in terms of a brane configuration
which is T-dual to ours. We will discuss these issues in more detail in section 3.

Let us now obtain the Seiberg-Witten solution for these ${\cal N}=2$ theories.
We start with the theory in Figure 1.
The method was explained in detail in~\cite{K}. We compactify $x^3$ on a circle
of radius $R$ and perform the $3-10$ flip. The brane configuration now consists
of $2N$ D4 branes wrapping ${\bf T}^2/{\bf Z}_2$ and $k$ D4$'$ branes
piercing them at $k$ points $p_1,\ldots,p_k$. It is more convenient to think
of D4 branes as wrapped on the double cover ${\bf T}^2$, with $2k$ punctures
located symmetrically with respect to a fixed point of the involution. We will
use an affine parameter $z$ as a coordinate on ${\bf T}^2$.
The theory on D4 branes is an impurity theory~\cite{KS} whose Higgs branch is the mirror of 
the Coulomb branch of the original theory compactified on a circle.
This Higgs branch is the moduli space of $U(2N)$ Hitchin equations with residues~\cite{KS}
\begin{eqnarray}\label{Hitchin}
&&F_{z\ov{z}}-[\Phi_z,\Phi_z^\dagger]=0, \nn \\
&&\ov{D}\Phi_z=-\frac{\pi}{RR_6}\sum_{\al=1}^{k}
(\delta^2 (z-z_\al)+\delta^2(z+z_\al))\ \diag (m_\al,0,\ldots,0).
\end{eqnarray}
The parameters $m_\al$ are related to the masses of the bifundamentals. Taking the
trace of the second equation in Eq.~(\ref{Hitchin}) one can see that
$m_\al$ satisfy $\sum_\al m_\al=0$. 
The Higgs field $\Phi$ describes the positions of the D4 branes in $x^4,x^5$, so
it has to satisfy
\be\label{proj2}
\Phi(z)=-M\Phi(-z)M^{-1}.
\ee
Here $M$ describes the action of the orbifold group on the Chan-Paton factors of the D4
branes. According to~\cite{DM} (see also~\cite{GLY}) $M$ generates a regular 
representation of ${\bf Z}_2$~\cite{DM}:
\be\nn
M=\left(\begin{array}{cc} 1_{N\times N} & 0 \\
                          0 & -1_{N\times N} 
        \end{array}\right)
\ee
Similarly, the $U(2N)$ gauge field $A_z$ must satisfy
\be\nn
A_z(z)=-MA_z(-z)M^{-1}.
\ee
The moduli space of solutions of Hitchin equations with these constraints has
a natural hyperk\"ahler metric. This metric is the exact metric on the Coulomb
branch of the $d=4$ $\CN=2$ theory compactified on a circle of radius $R$.
It is a very complicated metric, but things get simpler in the decompactification 
limit $R\ra\infty$. In this limit one is interested in the Seiberg-Witten curve, which
is the spectral cover of the Hitchin system~\cite{DW,K}. The spectral cover can
be constructed without actually solving Hitchin equations. By definition, it
is a $2N$-fold cover of the elliptic curve $\Sigma={\bf T}^2$ on which the Hitchin
system lives given by the equation
\be\label{spectral}
\det(v-\Phi(z))=0.
\ee
Explicitly, the spectral cover is given by
\be\label{curve}
v^{2N}+f_1 v^{2N-1}+f_2 v^{2N-2}+\cdots +f_{2N}=0,
\ee
where $f_1,\ldots,f_{2N}$ are meromorphic functions on $\Sigma$, by virtue of Hitchin
equations. They have simple poles at the points
$z_1,\ldots,z_k,-z_1,\ldots,-z_k$.
Furthermore, Eq.~(\ref{proj2}) implies that $f_{2j}$ are even functions of
$z$, while $f_{2j-1}$ are odd. Let us represent $\Sigma$ as a cubic curve
\be\label{torus}
y^2=(x-e_1)(x-e_2)(x-e_3).
\ee
The above conditions constrain $f_\ell$ to be of the form
\begin{eqnarray}\label{fi}
f_{2\ell}=\sum_{\al=1}^{k} \frac{a_{\ell\al}}{x-x_\al}+a_{\ell 0},\\ \nn
f_{2\ell-1}=\sum_{\al=1}^{k} \frac{y b_{\ell\al}}{x-x_\al},\quad \sum_\al b_{\ell\al}=0,
\end{eqnarray}
where $a_{\ell 0},\ldots,a_{\ell k}$ and $b_{\ell 1},\ldots, b_{\ell k}$ are complex 
constants, $x_\al={\cal P}(z_\al)$, and ${\cal P}(z)$ is the Weierstrass elliptic function.
Thus the spectral cover depends on $2kN$ parameters. Some of them are the
Coulomb branch moduli, and some are the parameters of the theory, namely the masses
of the bifundamentals $m_\al$. As in~\cite{Witten}, the mass parameters determine the
asymptotic behaviour of the noncompact curve $\Sigma$. Therefore they must be identified 
with the residues of $f_1$. Alternatively, it is easy to see from Hitchin equations that
the residues of $f_1$ are proportional to $m_\al$. Since $\sum_\al m_\al=0$,
we have $k-1$ mass parameters. All the other parameters of the curve are
moduli. Their number, $2kN-k+1$, agrees with the dimension of the Coulomb branch
expected from field theory.

Let us now turn to the brane configurations in Figures 2 and 3. The only
difference compared to the previous case is that the Higgs field is allowed to have
residues at some of the fixed points of $\Sigma$. For the theory in Figure 2 the 
Seiberg-Witten
curve is given by the same  equations Eqs.~(\ref{curve},\ref{torus},\ref{fi}),
except that now the parameters $b_{\ell\al}$ need not satisfy the constraint
$\sum_\al b_{\ell\al}=0$. The total number of parameters is $(2k+1)N$, out of which
$k$ are mass parameters and $2kN+N-k$ are moduli, in agreement with field
theory.
For the theory in Figure 3 the functions $f_\ell$ are given by
\begin{eqnarray}
f_{2\ell}=\sum_{\al=1}^{k} \frac{a_{\ell\al}}{x-x_\al}+a_{\ell 0},\nn \\ \nn
f_{2\ell-1}=\sum_{\al=1}^{k} \frac{y b_{\ell\al} }{x-x_\al}+\frac{yb_{\ell 0}}{x-e_3}.
\end{eqnarray}
The number of moduli is $(2N-1)(k+1)$, while the number of mass parameters is
$k+1$. This agrees with the field theory count.

We mentioned above that the restriction on the masses of the bifundamentals 
$\sum_\al m_\al =0$ can be removed by introducing a shift along $x^4,x^5$ as 
one goes around $x^6$~\cite{Witten}.
A way of introducing this deformation into the Hitchin equations was explained
in~\cite{K}: one simply replaces ${\rm diag}(m_\al,0,\ldots,0)$ in Eq.~(\ref{Hitchin})
with ${\rm diag}(m_\al,-M,\ldots,-M)$. Then one can easily see that 
the constraint becomes $\sum_\al m_\al =(2N-1)kM$. We will not discuss the 
corresponding modification of the Seiberg-Witten curve, but we will keep in mind that
the total number of mass parameters is $k$, $k+1$, and $k+2$ for the theories in
Figures 1, 2, and 3, respectively. 

Given Seiberg-Witten curves it is easy to derive the S-duality group of the
theories. In all three cases the gauge couplings and theta-angles are determined
by the $\tau$-parameter of $\Sigma$ and the location of the punctures. Let us 
denote by ${\cal M}_{p,k}$ the moduli space of an orbifolded elliptic curve
$\Sigma/{\bf Z}_2$,
with $k$ marked points not coinciding with the orbifold points, and with $p$
orbifold points marked as well ($p=0,1,2$). Then the S-duality groups of the
theories in
Figures 1,2, and 3 are the fundamental groups of ${\cal M}_{0,k}$,
${\cal M}_{1,k}$, and ${\cal M}_{2,k}$, respectively.

If we set all mass parameters to zero and go to the origin of the moduli space,
we obtain an ${\cal N}=2$ superconformal field theory. It has a number of 
deformations which preserve supersymmetry, and from the curve we can 
read off their R-charges and dimensions. Since the deformation parameters 
are chiral primary fields which are $SU(2)_R$ singlets, 
their dimensions and R-charges are related by $\Delta=R/2$~\cite{SWAP}.

First, there are exactly marginal deformations obtained by varying the
gauge couplings and theta-angles of the theory. They have zero R-charge and
dimension. They are encoded in the locations of the punctures and the complex
structure of $\Sigma$. Their total number is $k+1$. 
Second, there are Coulomb branch moduli and the masses of the
bifundamentals. The R-charge of the moduli can be determined from the curve as
follows. The R-symmetry is realized in the brane configuration as a rotation
in the
$x^4,x^5$ plane, and the standard normalization is such that the Higgs field has
R-charge 2. This means that $v$ in Eq.~(\ref{curve}) has R-charge 2. As for
$x$ and $y$, they have zero R-charge. This determines the R-charges of all
parameters in Eq.~(\ref{curve}). Namely, $a_{\ell\al}$ has R-charge $4\ell$, and
$b_{\ell\al}$ has R-charge $4\ell-2$. In particular the masses (i.e. $b_{1\al}$) have
R-charge 2. This is the expected result: the masses are the lowest components of
background vector multiplets  which couple to conserved currents, therefore their
R-charge is 2 in any ${\cal N}=2$ theory. Since the deformation parameters 
are chiral primary fields which are $SU(2)_R$ singlets, 
their dimensions and R-charges are related by $\Delta=R/2$~\cite{SWAP}.

As mentioned above, it is impossible to introduce masses for fundamental
hypermultiplets without making the M-theory background curved. Our solution is
only valid when these masses are zero. Still, we know on general grounds that
these deformations exist, and their R-charge is 2.

\subsection{${\bf Z}_n$ orbifolds for $n>2$}

We now consider M-theory on the orbifold $({\bf T}^2\times {\bf R}^2)/{\bf Z}_n$  for
$n=3,4,6$. The $\tau$-parameter of the torus is $\exp(i \pi/3)$ or $i$ depending
on whether $n=3,6$ or $n=4$. The configuration also includes $k$ M5$'$
wrapping the base of this fibration and $N$ M5 branes wrapping the fiber. These
configurations do not have a IIA reduction. Nevertheless we can still find the Seiberg-Witten
solution for the theory on the M5 branes using the method of~\cite{K}, i.e.
by compactifying $x^3$ and performing the $3-10$ flip. We obtain $nN$ D4 branes
wrapping ${\bf T}^2/{\bf Z}_n$ with $k$ punctures. On the $n$-fold cover of
${\bf T}^2$ we thus have $nk$ punctures located at
$z=\omega^j z_\al, \al=1,\ldots,k, j=0,\ldots,n-1$. Here 
$\omega=\exp(2\pi i/n)$.
The Higgs branch of the corresponding impurity theory is described by $U(nN)$ Hitchin
equations of the form
\begin{eqnarray}\label{Hitp}
&&F_{z\ov{z}}-[\Phi_z,\Phi_z^\dagger]=0, \nn \\
&&\ov{D}\Phi_z=-\frac{\pi}{RR_6}\sum_{\al=1}^{k} \diag (m_\al,0,\ldots,0)
\sum_{j=0}^{n-1}\delta^2 (z-\omega^j z_\al).
\end{eqnarray}
The Higgs field and the gauge connection are in the adjoint of $U(nN)$. They have
to satisfy
\be\label{projn}
\Phi(\omega z)=\omega^{-1} M\Phi(z) M^{-1},\quad A_z(\omega z)=\omega^{-1}M A_z(z)M^{-1},
\ee
where 
\be\nn
M=1_{N\times N}\otimes{\rm diag}(1,\omega,\ldots,\omega^{n-1})
\ee
generates the regular representation of ${\bf Z}_n$. The trace of the second 
equation in Eq.~(\ref{Hitp}) implies $\sum_\al m_\al=0$. As before, this condition can be
relaxed by introducing an analogue of the global mass, namely by replacing
$\diag (m_\al,0,\ldots,0)$ in Eq.~(\ref{Hitp}) by $\diag (m_\al,-M,\ldots,-M)$.
Then $m_\al$ satisfy $\sum m_\al=(nN-1)kM$. We do not consider this modification in what
follows.

Another interesting modification is to introduce punctures at the fixed points of 
the orbifolded torus. This amounts to having residues for $\Phi$ at the fixed points. 
It is easy to
see that this is consistent with the projection Eq.~(\ref{projn}). Recall that
a puncture away from the fixed point together with its $n-1$ images corresponds to 
an M5$'$ brane wrapping the base ${\bf R}^2/{\bf Z}_n$.
Then a puncture sitting at a fixed point of the orbifold must correspond to 
a ${1/n}^{\rm th}$ of the usual M5$'$ brane. Although this is an interesting possibility,
we will not consider it here.

The Hitchin system defined above provides a solution for the probe theories
compactified on a circle of radius $R$. The Seiberg-Witten curve is the spectral 
cover of the Hitchin system. It has the form
\be\label{curven}
v^{nN}+f_1 v^{nN-1}+f_2 v^{nN-2}+\cdots+f_{nN}=0.
\ee
The functions $f_\ell$ are meromorphic functions with simple poles at $nk$ points
$z=\omega^j z_\alpha$. By virtue of Eq.~(\ref{projn}) they satisfy
\be\label{projfn}
f_\ell(\omega z)=\omega^{-\ell} f_\ell(z).
\ee
These conditions completely determine the Seiberg-Witten curve.

As an example, let us work out the explicit form of the curve for $n=3$. It will
be obvious then how to extend the discussion to larger $n$. The elliptic curve
with a ${\bf Z}_3$ automorphism can be thought of as a cubic curve $y^2=x^3-1$.
The solution of Eq.~(\ref{projfn}) looks as follows:
\be\label{fi3}
f_\ell=\sum_{\al=1}^k a_{\ell\al}(y+y_\al) \sum_{j=0}^2
\frac{\omega^{-\ell j}}{\omega^{2j}x-x_\al} + b_\ell \sum_{j=0}^2
\omega^{\ell j}.
\ee
Here $x_\al={\cal P}(z_\al), y_\al={\cal P}'(z_\al)/2$, $a_{\ell\al}$ and $b_\ell$
are complex
constants. Note that the last term in Eq.~(\ref{fi3}) is nonvanishing only if
$\ell=0\ {\rm mod}\ 3$. In addition the coefficients $a_{\ell\al}$ must satisfy
a constraint which follows from the requirement that $f_\ell$'s be nonsingular at
$x=\infty$:
$$\sum_{\al=1}^k a_{\ell\al}\sum_{j=0}^2 \omega^{-j(\ell+2)}=0.$$
This constraint is nontrivial only if $\ell=1\ {\rm mod}\ 3$. Now we can count the
number of parameters in the equation of the curve. Let us denote the dimension of the space
of $f_\ell$'s satisfying Eq.~(\ref{projfn}) by $d_\ell$. Then we get
\be\label{nmod}
d_\ell=\left\{ \begin{array}{ll}
k+1, & \ell=0 \ {\rm mod}\ 3 \\
k-1, & \ell=1 \ {\rm mod}\ 3 \\
k, & \ell=2 \ {\rm mod}\ 3 \end{array} \right.
\ee
It follows that the total number of parameters is $3kN$.
The R-charges of the parameters can be read off the curve in the same
way as in the previous subsection: the parameters entering $f_\ell$ have
R-charge $2\ell$.
Since we have no Lagrangian description of the theory, we have nothing to
compare these results with. Nevertheless, we can
determine which of these parameters are moduli and which are ``masses.''
The $k-1$ residues of $f_1$ determine the asymptotic behaviour of the curve,
so they are the ``masses.'' Alternatively, it is easy to see from Hitchin equations
that these residues are proportional to $m_\al$'s which are the parameters of the
system. Their R-charge is $2$, as is appropriate for mass parameters.
They may be regarded as living in vector multiplets. (It is on these grounds
that we call them masses.) One should not forget about the possibility of introducing
the global mass $M$ (see above). Taking it into account, the total number of 
background vector multiplets is $k$. The possibility to couple $k$ vector multiplets
implies that the theory in question has $k$ conserved $U(1)$ currents. Of course, the
actual current algebra may be bigger than that. Recall that our curve for the 
${\bf Z}_2$ orbifold did not contain masses for the fundamentals and therefore missed
flavor symmetries acting on them. It seems likely that the same
is true for other orbifolds. In the next section we will discuss this question from the
point of view of F-theory. It will be shown that for $n=2$ and $k=0$ (no punctures) 
the theory has a $E_6$ current algebra (in addition to the $U(1)$'s found above),
while for $k>0$ it has some subalgebra of $E_6$.

Similar analysis can be performed for $n=4$ and $n=6$. Here we just
state the results. For all $n$ we find
\be\label{dimdj}
d_\ell=\left\{ \begin{array}{ll}
k+1, & \ell=0 \ {\rm mod}\  n \\
k-1, & \ell=1 \ {\rm mod}\  n \\
k & {\rm otherwise} \end{array} \right.
\ee
The total number of parameters in the equation of the curve is $nkN$.
Out of these $k$ are ``masses'' (including the global mass) and the rest are moduli. 
Thus for any $n$ we have $k$ $U(1)$ currents.
It will be shown in the next section that for $k=0$, in addition to these $U(1)$ currents, 
the $n=4$ and $n=6$ theories have $E_7$ and $E_8$ current algebras, respectively.
For $k>0$ certain subalgebras of these current algebras remain.

Finally, let us determine the S-duality groups. We have $k$ exactly marginal
deformations related to the location of the punctures. Therefore the S-duality
group is the fundamental group of the moduli space of an orbifolded
elliptic curve $\Sigma/{\bf Z}_n$ with $k$ marked points.

\section{F-theory duals}

In the limit when the volume of the elliptic fiber goes to zero, M-theory on
$({\bf T}^2\times {\bf R}^2)/{\bf Z}_n$ is equivalent to F-theory on the same
manifold. This manifold can be regarded as an orbifold limit of a
``noncompact K3.''
Since the $\tau$-parameter of ${\bf T}^2$ is constant, we get an F-theory
background with constant coupling \cite{sen,DasMu}. For $n=2$ this background is nothing but
a IIB orientifold background ${\bf R}^2/(\Omega (-1)^{F_L}{\cal R}_{45})$
with four D7 branes on top of the fixed point. Here ${\cal R}_{45}$ is the 
reflection of $x^4,x^5$. (We could also obtain this result by first
taking a IIA limit of the M-theory configuration and then T-dualizing along
$x^6$.) For $n=3,4,6$ the F-theory backgrounds can be thought of as a collection of
several mutually nonlocal 7-branes in IIB. In all four cases the theory
living on the worldvolume of the 7-branes has nonabelian gauge group $G$. For $n=2,3,4,6$
this group is $Spin(8)/{\bf Z}_2,E_6,E_7,E_8$, respectively.

The M-theory background studied in the previous section also contained $nk$ M5$'$ branes
wrapping the base ${\bf R}^2$ and $nN$ probe M5 branes wrapping ${\bf T}^2$. 
Let us first consider the case $k=0$, when M5$'$ branes
are absent. The probe M5 branes turn into
$N$ D3 branes parallel to $x^0,x^1,x^2,x^3$. For $n=2$ the theory on D3 branes
is an ${\cal N}=2$ $Sp(N)$ gauge theory with an antisymmetric tensor and four
fundamentals, in agreement with IIA arguments~\cite{probe}. 
For $n=3,4,6$ the probe theory does not have
a Lagrangian description. For any $n$ the 7-brane gauge symmetry $G$ is
the global symmetry of the probe theory. Thus for $n=3,4,6$ the probe theory
has an $E_6,E_7,E_8$ current algebra, respectively. This justifies the claims made
in the previous section.

Now let us consider the case $k>0$. In M-theory we have $nk$ M5$'$
branes, counting images. In F-theory they become an orbifold ${\bf C}^2/{\bf
Z}_{nk}$. Here ${\bf C}^2$ has coordinates $x^6,x^7,x^8,x^9$, and therefore the orbifold
plane is parallel to $x^0,x^1,x^2,x^3,x^4,x^5$. Thus the full F-theory
background is $({\bf T}^2\times {\bf R}^2)/{\bf Z}_n\times{\bf C}^2/{\bf Z}_{nk}$. It
can be thought of as an orbifold limit of K3$\times$K3 near the orbifold singularity.

The blow-up modes of the ${\bf C}^2/{\bf Z}_{nk}$ orbifold
correspond to the motion of M5$'$ branes along $x^7,x^8,x^9$. However there are only $k$ 
such modes, since an M5$'$ brane and its $n-1$ images move together. This means that
the ${\bf Z}_{nk}$ singularity can only be resolved to a product of $k$
${\bf Z}_n$ singularities, most of the blow-up modes being ``frozen.'' We may call this
branch of vacua the Coulomb branch. Instead of the blow-up modes one has $k$ extra complex
scalars which are the lowest components of $d=4$ $\CN=2$ vector multiplets. Each
such scalar corresponds to the $x^6,x^{10}$ position of an M5$'$ brane on the M-theory
side.
As in the ${\bf Z}_2$ case, the blow-up modes can be ``thawed'' by passing to a 
Higgs branch. To this end one must tune the above mentioned complex scalar to a 
particular value. At this point the missing
$n-1$ blow-up modes become massless, and one can give them VEVs. On the M-theory
side this corresponds to bringing an M5$'$ brane and its $n-1$ images to a fixed point
of the orbifold and then separating them in the $x^7,x^8,x^9$. 

Because of the freezing of the blow-up modes, 
the F-theory background is not really a geometric orbifold
$({\bf T}^2\times {\bf R}^2)/{\bf Z}_n\times{\bf C}^2/{\bf Z}_{nk}$.
A similar phenomenon occurs in Type I theory on 
${\bf C}^2/{\bf Z}_2$. It was shown by Polchinski~\cite{P} that there are two
different orientifolding procedures for Type IIB on ${\bf C}^2/{\bf Z}_2$.
The first procedure is simply to quotient by worldsheet parity $\Omega$. 
Then, as is common in perturbative constructions, there is a trapped flux
of the B-field through the shrunk 2-cycle, implying that the $Spin(32)/{\bf Z}_2$
bundle on the 9-branes has a nontrivial generalized Stiefel-Whitney class 
$\tilde{w}_2$~\cite{six,SS}. The 2-cycle can be blown up, and then one discovers a 
$Spin(32)/{\bf Z}_2$ instanton sitting where the fixed point used to be~\cite{six}.
This instanton breaks $Spin(32)/{\bf Z}_2$ down to $U(16)/{\bf Z}_2$.
This type of ${\bf Z}_2$ singularity arises in the Gimon-Polchinski model~\cite{GP}.

The second procedure is to quotient by $\Omega J$. Here $J$ is the symmetry of
the worldsheet conformal field theory of ${\bf C}^2/{\bf Z}_2$ which flips the 
sign of the twisted sector.
The projection $\Omega J$ kills the zero mode responsible for the blow-up,
keeping instead a $(1,0)$ tensor multiplet. Also, it does not have a trapped
flux of the B-field. As in the previous case tadpole cancellation requires 16 9-branes
carrying an $Spin(32)/{\bf Z}_2$ bundle. It also requires this bundle to have a 
nontrivial monodromy $\CM$ when the singular point of ${\bf C}^2/{\bf Z}_2$ is deleted. 
($\CM$ is encoded in a notrivial action of the orbifold group on the
Chan-Paton factors). The monodromy breaks $Spin(32)/{\bf Z}_2$ down to
$(Spin(16)\times Spin(16))/{\bf Z}_2$. This second orientifold is not a geometric
orbifold of Type I, because it lacks a blow-up mode. However, one can pass to a 
geometric phase by tuning the real scalar in the
tensor multiplet, so that tensionless strings arise. At this point one can go to the Higgs branch,
where the tensor multiplet is lifted, and the singularity is resolved.
This transition was described in detail in~\cite{IB,HZ}. Unlike the Coulomb branch
(i.e. the branch with the frozen ${\bf Z}_2$ singularity),
the Higgs branch cannot be realized by a free CFT. 

Guided by these Type I considerations one can easily guess the right
orbifold action in our case. In fact, as remarked in section 2, for $n=2$ our
brane configuration is T-dual to that in~\cite{HZ}, and therefore also to that
in~\cite{P}. It follows that one can obtain the right orbifold action simply by
T-dualizing the construction in~\cite{P} along two directions parallel to the orbifold
plane ($x^4,x^5$ in our notation). It is also easy to guess how to generalize
to $n\neq 2$. One should start with F-theory on 
$${\bf T}^2\times {\bf R}^2\times {\bf C}^2/{\bf Z}_{nk},$$
where the complex structure of ${\bf T}^2$ is fixed appropriately.
Then one should orbifold further by $\CP_n J_n$, where
$\CP_n$ is the generator of the SUSY-preserving ${\bf Z}_n$ action on 
${\bf T}^2 \times {\bf R}^2$, and $J_n$ multiplies the twisted sectors of the
${\bf Z}_{nk}$ orbifold by phases. It is the presence of $J_n$ which distinguishes
this construction from a geometric orbifold. To describe $J_n$ precisely it is 
convenient to think of $nk-1$ twisted sectors as arising from $nk-1$ shrunk 
2-cycles. They are acted upon by a regular representation of ${\bf Z}_{nk}$. One 
can choose the basis of the 2-cycles so that the generator of ${\bf Z}_{nk}$ 
multiplies the $s^{\rm th}$ sector by $\exp(2\pi is/nk)$. In this basis $J_n$ 
acts by multiplying the $s^{\rm th}$ sector by $\omega^s$, where 
$\omega=\exp(2\pi i/n)$. In other
words, $J_n$ generates a natural ${\bf Z}_n$ subgroup of ${\bf Z}_{nk}$. For $n=2$
it is easy to see that this construction is T-dual to that in~\cite{P} and
produces frozen ${\bf C}^2/{\bf Z}_2$ singularities ($k$ of them). It is natural to
expect that in general one gets $k$ frozen ${\bf C}^2/{\bf Z}_n$ singularities, 
because the above ``orientifolding'' procedure makes sense only when ${\bf Z}_n$
singularities are present, and not on a resolved ALE space.

Next we have to figure out the $d=4$ gauge group $H$ due 7-branes. It becomes
part of the global symmetry of the D3 probes. For $k=0$ 
$H$ is the same as the eight-dimensional gauge group, i.e. $H=Spin(8)/{\bf Z}_2,E_6,E_7,$ 
or $E_8$, depending on the value of $n$.
But for $k>0$ the vector bundle on the 7-branes can have
a nontrivial monodromy $\CM$ when the singular point of ${\bf C}^2/{\bf Z}_{nk}$
is removed. $\CM$ breaks the eight-dimensional gauge group $G$ down to a subgroup $H\subset G$.
We can be slightly more specific about $H$. Since the fundamental group
of ${\bf C}^2/{\bf Z}_{nk}$ with the origin removed is ${\bf Z}_{nk}$, 
$\CM$ generates at most a ${\bf Z}_{nk}$ subgroup  
In fact, we expect $\CM$ to generate a ${\bf Z}_n$ subgroup, since the ${\bf Z}_{nk}$ 
singularity can be resolved to a product of $k$ ${\bf Z}_n$ singularities
with identical monodromies. Unfortunately there are many inequivalent choices of $\CM$ 
resulting in different $H$. Thus additional constraints are required. 

The problem of computing $H$ is part of a more general problem, namely
understanding the twisted sectors of the $\CP_n J_n$ projection. For $n=2$
the twisted sectors are simply open strings, and their choice is constrained
by perturbative consistency conditions of~\cite{GP,P}. These conditions 
uniquely fix $H=(Spin(4)\times Spin(4))/{\bf Z}_2$. This is most easily
seen by T-dualizing the set-up of~\cite{P}. The monodromy is realized as
the action of the ${\bf Z}_2$ orbifold group on the Chan-Paton factors.
Tadpole cancellation requires $\CM$ to be of the form
\be\label{monodr}
\CM=\left(\begin{array}{cc} 1_{4\times 4} & 0 \\
                        0 & -1_{4\times 4}\\
                        \end{array}\right).
                        \ee
Clearly, it breaks $Spin(8)/{\bf Z}_2$ down to $(Spin(4)\times Spin(4))/{\bf Z}_2$. 
Not suprisingly, $H$ agrees with the flavor symmetry of the gauge theory on the
probes branes (see section 2).

For $n>2$ twisted states include multi-prong strings~\cite{GZw}. It is far from
clear what replaces the conditions of~\cite{GP,P} in this case. It would
be very interesting to pin down the exact content of the twisted sector
of the F-theory orbifold which corresponds to the M-theory configuration of section 2. 
In any case, twisted states must include a vector multiplet in the adjoint of $H$.  

A peculiarity of the $n=2$ case is that the eight-dimensional gauge group 
$G=Spin(8)/{\bf Z}_2$ is not simply connected. Thus one may also consider a 
${\bf Z}_2$ orbifold with a nontrivial generalized Stiefel-Whitney class. Such a 
${\bf Z}_2$ singularity is T-dual to that constructed by Gimon and Polchinski~\cite{GP} 
and therefore can be blown up. The corresponding M-theory configuration has 
two unpaired NS5 brane at the orientifold planes, as in Figure 3. 
For $n>2$ the gauge group is simply connected,
so no analogue of the Stiefel-Whitney class exists. Still, one can have a configuration
analogous to Figure 3, namely $n$ M5$'$ branes stuck at $n$ fixed points of the
${\bf Z}_n$ orbifold. It is not clear to us what this configuration maps to 
in F-theory. In what follows we will not consider F-theory configurations 
corresponding to M-theory backgrounds with stuck M5$'$ branes.

\section{The large $N$ limit}

When the number $N$ of three-branes at the singularity 
$({\bf T}^2\times {\bf R}^2)/{\bf Z}_n\times{\bf C}^2/{\bf Z}_{nk}$
becomes large, $\CN=2$ superconformal field theory
on the brane world-volume is expected to be "holographic"
to the near horizon limit of a supergravity solution \cite{maldacena}. 
In our case this near horizon limit is $AdS^5\times S^5/({\bf Z}_{nk}\times {\bf Z}_n)$.
To describe the ${\bf Z}_{nk}\times {\bf Z}_n$ action we consider the 
$SU(2)_L\times SU(2)_R\times U(1)_R$ subgroup of $SU(4)$ which is a cover of the 
isometry group of $S^5$.
${\bf Z}_{nk}$ is embedded in the $SU(2)_L$ factor. The generator of
${\bf Z}_n$ looks like $\CP_n J_n$, where $\CP_n$ acts as a ${\bf Z}_n$ automorphism 
on the ${\bf T}^2$ of
F-theory and simultaneously as a $U(1)_R$ rotation, while $J_n$ multiplies the
twisted sectors of the ${\bf Z}_{nk}$ orbifold by phases, as described in section 3.
This orbifolding reduces supersymmetry from sixteen to eight supercharges
and breaks $SU(4)$ down to $SU(2)_R\times U(1)_R \times U(1)_L$. The $SU(2)_R\times
U(1)_R$ becomes the R-symmetry of the SCFT on the boundary, while $U(1)_L$
becomes a global non-R symmetry. We will identify this $U(1)_L$ symmetry more
precisely later.

According to the AdS/SCFT correspondence~\cite{AdS} conformal dimensions of 
operators in the boundary SCFT are related to masses of the supergravity states. For a p-form
state this relation is
\be\label{scdim}
m^2 = (\Delta-p) (\Delta+p-4).
\label{dim}
\ee
In this section we compute the supergravity spectrum and compare it
with the results of section 2. We consider only states which couple to
relevant and marginal operators.
We will show that there is complete agreement 
between the AdS approach and the the approach based on the analysis of the 
Seiberg-Witten curve. 

Before plunging into computational details
let us make a few general comments. Due to supersymmetry
it is sufficient to consider only bosonic states. 
Supersymmetry also protects the dimensions of the operators
and the masses of the supergravity modes from $\alpha'$ corrections as
long as they come in short multiplets. In particular, the
dimension of a chiral primary operator is determined solely by its
$U(1)_R$ charge and the dimension $d$ of its $SU(2)_R$
representation:
\be
\Delta = \left| {R \over 2} \right|+d-1.
\label{unit}
\ee
One can divide supergravity states into three different
groups according to their origin.
First, there are states coming from projecting the supergravity spectrum on 
$AdS^5\times S^5$. We will call them bulk modes. Second, there are 
twisted states of the ${\bf Z}_{nk}$ orbifold which are invariant with respect to
$\CP_n J_n$.
Third, there are states twisted with respect to $\CP_n J_n$. They are charged 
with respect to the 7-brane gauge group $H$. 
We consider these three groups of states in turn.

\subsection{Bulk supergravity states}

The spectrum of the bulk modes can be obtained by decomposing $SU(4)$
representations of supergravity on $AdS_5 \times S^5$ into
representations of $SU(2)_L \times SU(2)_R \times U(1)_R$,
and projecting onto states invariant under the orbifold action.
The ${\bf Z}_{nk}$ action is embedded into the $SU(2)_L$ factor, while
the geometric part of the $\CP_n$ action is embedded into $U(1)_R$.
The commutant of ${\bf Z}_{nk}$ in $SU(2)_L$ is
the $U(1)_L$ symmetry corresponding to a non-R symmetry on the boundary.
The projection on the invariant states proceeds in two steps.
In the first step we project on the ${\bf Z}_{nk}$-invariant states. 
The states with zero $U(1)_L$ charge are automatically ${\bf Z}_{nk}$ invariant. 
Every odd-dimensional representation of $SU(2)_L$ contains exactly one such state,
while even dimensional representations contain none~\cite{toz}. More generally,
a state is ${\bf Z}_{nk}$ invariant if and only if its $U(1)_L$ charge is a
multiple of $nk$. Both even-dimensional and odd-dimensional representations of
$SU(2)_L$ may contain such states. In the second
step we require that the states be invariant with respect to $\CP_n J_n$.
For states which originate from the $d=10$ fields other than the $B$-fields this
simply means that the R-charge is a multiple of $2n$~\cite{fs,afm}.
Special attention should be paid to the modes coming
from the reduction of the $B$-fields, since the nongeometric part of $\CP_n$ 
induces non-trivial monodromy on these fields~\cite{fs}.
After one diagonalizes the monodromy matrix one finds that the selection
rule for these modes is $R=\pm 2$ mod $2n$. It is sufficient to consider only
the plus sign, since the other choice of sign is obtained by complex conjugation.

The supergravity spectrum on $AdS^5\times S^5$ has been worked out in~\cite{krn}.
Supergravity states (and hence the SCFT operators) organize themselves into
multiplets of $SU(2,2|2)$. Below we discuss bosonic states which survive
the orbifold projection, limiting ourselves to those states which couple to relevant
or marginal operators. We first discuss the states
with zero $U(1)_L$ charge, and then make some comments about states which transform
nontrivially under $U(1)_L$.

The graviton is a singlet of $SU(4)$ and
therefore is in the spectrum of any ${\bf Z}_{nk} \times {\bf Z}_n$ orbifold.
It couples to the stress-energy tensor (marginal operator of
dimension $4$). The spectrum of supergravity on $AdS^5\times S^5$ also contains a 
massless gauge boson in the adjoint ({\bf 15})
of $SU(4)$. A decomposition of {\bf 15} into irreps of 
$SU(2)_L\times SU(2)_R\times U(1)_R$ yields {\bf 15} = {\bf (1,1)}$_0$ + {\bf (2,2)}$_2$ +
{\bf (2,2)}$_{-2}$ + {\bf (3,1)}$_0$ + {\bf (1,3)}$_0$.
The projection on the ${\bf Z}_{nk}$-invariant states leaves
{\bf 1}$_0$ + {\bf 1}$_0$ + {\bf 3}$_0$, which is automatically
${\bf Z}_n$-invariant. The ${\bf 3}_0$ and one of the ${\bf 1}_0$ 
gauge bosons couple to the $SU(2)_R$ and $U(1)_R$ currents on the boundary, 
respectively. The remaining massless vector is a $U(1)_L$ gauge boson which couples to a 
non-R current on the boundary. In section 2 we found that the
boundary SCFT has $k$ $U(1)$ currents. The current coupled to the $U(1)_L$ gauge boson is 
one of them. (We will make this identification
more precise below). The remaining $k-1$ currents of the SCFT must
couple to gauge bosons which come from the twisted sectors.

Now let us turn to 2-form states. They satisfy first-order equations of the
antiself-dual type (and their complex conjugates satisfy equations of the self-dual 
type)~\cite{krn,gun}. The ones which couple to relevant or marginal operators satisfy 
$m^2=1$ and $m^2=4$ and transform as {\bf 6} and {\bf 20}$'$ of $SU(4)$, respectively. 
These 2-forms originate from the B-fields in $d=10$, so the selection rule for their R-charges
is $R=2$ mod $2n$. {\bf 6} decomposes with respect to $SU(2)_L\times SU(2)_R\times U(1)_R$ as
{\bf (2,2)}$_0$+{\bf (1,1)}$_2$+{\bf (1,1)}$_{-2}$.
For $n>2$ only {\bf (1,1)}$_2$ survives, while for $n=2$ {\bf (1,1)}$_{-2}$ survives as well.
These states couple to operators of dimension 3 on the boundary which transform
as {\bf 1}$_{\mp 2}$. Recall that the graviton multiplet of $SU(2,2|2)$ contains
an antiself-dual tensor which lives in {\bf 1}$_2$ of $SU(2)_R\times U(1)_R$. Since
the graviton is present for all $n$, we must identify {\bf 1}$_2$ 2-form we found above
as the superpartner of the gravition. The other 2-form ({\bf 1}$_{-2}$) which is present
only for $n=2$ then must be a member of some other $SU(2,2|2)$ multiplet. Such
a multiplet is known as a tensor multiplet~\cite{gun}. Precisely for
$n=2$ the SCFT on the boundary admits a Lagrangian description, so we can
identify the operator {\bf 1}$_{-2}$ couples to as $\sum_\al\Tr (F_\al+i{\tilde F}_\al)\Phi_\al$. 
The notation here is as follows: $F_\al$ is the field strength of the $\al^{\rm th}$
factor in the gauge group, ${\tilde F}_\al$ is its dual, $\Phi_\al$ is the complex scalar 
in the $\al^{\rm th}$ vector multiplet. The sum over $\al$ arises because we are 
discussing the bulk sector. 

The remaining 2-form ({\bf 20}$'$) is projected out completely for $n=2,4,6$, 
while for $n=3$ a state which transforms as {\bf 1}$_{-4}$ remains. This 2-form couples
to an SCFT operator of dimension 4 which transforms as {\bf 1}$_4$ with respect
to $SU(2)_R\times U(1)_R$. Since we do not have a Lagrangian description of the SCFT
for $n=3$, we cannot write down an explicit form for this operator. However, we
will see later that this operator is a superpartner of a Coulomb branch
modulus appearing in the Seiberg-Witten curve. Thus its existence follows from
the results of section 2 combined with $\CN=2$ supersymmetry.

It remains to analyze the scalar modes. The scalars which couple to marginal or relevant 
operators have $m^2\leq 0$ and come in representations {\bf 1}$_C$, {\bf 20}$'$, 
{\bf 10}$_C$, {\bf 50}, {\bf 45}$_C$, and {\bf 105} of $SU(4)$.
The easiest to deal with is the massless complex
scalar {\bf 1}$_C$ coming from the reduction of the IIB dilaton.
It is a singlet of $SU(4)$, so one could make a hasty conclusion that it remains
in the orbifolded spectrum. However, one should not forget that $\CP_n$ includes
a fractional-linear transformation of the IIB dilaton. For $n=2$ this is an
identity transformation, so {\bf 1}$_C$ indeed survives. But for $n>2$ invariance
with respect to $\CP_n$ requires it to be a root of unity, so {\bf 1}$_C$ is
projected out. This is in perfect agreement with the spectrum of 2-forms found
above. Indeed, for $n=2$ we found an antiself-dual 2-form {\bf 1}$_{-2}$ which we
interpreted as the highest component of the tensor multiplet of $SU(2,2|2)$ with
$m^2=1$. The highest component of such a multiplet is a massless complex scalar in 
{\bf 1}$_0$. Thus $SU(2,2|2)$ invariance requires a massless {\bf 1}$_0$ state. 
On the other hand for $n>2$ the 2-form {\bf 1}$_{-2}$ is projected out, so we do not 
expect to have any massless {\bf 1}$_0$ scalar.

The analysis of the remaining scalars is straightforward. The results are summarized
in Tables 1---4. To make tables shorter we only listed the scalars which are
primary with respect to $SU(2,2|2)$. In most cases we indicated the type of $SU(2,2|2)$ 
multiplets they belong to. The graviton multiplet ($G$) contains a {\bf 1}$_0$ scalar 
with $m^2=-4$, four massless vectors in the adjoint of $SU(2)_R\times U(1)_R$,
a {\bf 1}$_2$ 2-form with $m^2=1$ satisfying an equation of the antiself-dual type,
and a graviton. (We already saw all the states in this multiplet except the scalar,
which arises from {\bf 20}$'$). The Maxwell multiplet ($V$) contains a real scalar in
{\bf 3}$_0$ with $m^2=-4$, a complex scalar in {\bf 1}$_{-2}$ with $m^2=-3$,
and a massless {\bf 1}$_0$ vector. There is only one such multiplet in the bulk spectrum:
the one containing the $U(1)_L$ gauge boson discussed above. The antiself-dual
tensor multiplet $T_p$ contains a complex scalar in {\bf 1}$_{-2p}$ with $m^2=p(p-4)$,
a real scalar in {\bf 3}$_{-2p+2}$ with $m^2=(p+1)(p-3)$, a complex antiself-dual 2-form 
in {\bf 1}$_{-2p+2}$ with $m^2=(p-1)^2$, and a complex scalar in {\bf 1}$_{-2p+4}$
with $m^2=p^2-4$. The 2-forms we found above fit in $T_2$ and $T_3$, respectively. The tables
also contain $T_4$; the corresponding 2-form couples to an irrelevant ($\Delta=5$) operator 
and thus did not appear in our analysis. 

For $n=2$ the boundary CFT admits a Lagrangian description in terms of
an $Sp(N)\times SU(2N)^{k-1}\times Sp(N)$ gauge theory with two fundamentals for 
each $Sp(N)$ factor and $k$ bifundamentals.
In Table 1 we indicated which
gauge theory operators the supergravity states couple to. 
The notation is as follows: $\Phi_\al$ is a complex scalar in the $\CN=2$ vector multiplet,
$\tilde{Q}_\al, Q_\al$ are complex scalars in a bifundamental hypermultiplet. The index
$\al$ runs from 1 to $k$. Note
that we identified the operator coupled to the primary state in the Maxwell
multiplet as $\sum_\al {\tilde Q}_\al Q_\al$. This particular linear combination
of ${\tilde Q}_\al Q_\al$ is the only one invariant with respect to the ${\bf Z}_k$
subgroup of the orbifold group ${\bf Z}_{nk}$. The supermultiplet containing
$\sum_i {\tilde Q}_i Q_i$ also contains $\int d^2\theta \sum_i {\tilde Q}_i Q_i$
which couples to the global mass, and a $U(1)$ current. This $U(1)$ current
is the Noether current corresponding to the invariance of the theory
with respect to multiplying all $Q$'s by $e^{-i\phi}$ and all $\tilde Q$'s by $e^{i\phi}$.
It follows that the $U(1)_L$ symmetry must be identified with this symmetry of the
gauge theory.
\begin{table}\begin{center}
\begin{tabular}{|c|c|c|c|}
\hline
$SU(2,2|2)$ & $SU(2)_R \times U(1)_R$ & $\Delta$ & SCFT operator\\
\cline{1-4}
$G$ & {\bf 1}$_0$ & $2$ & \\
$V$ & {\bf 3}$_{0}$  & $2$ & $\sum_\al\Tr\tilde Q_\al Q_\al$ \\
$T_2$ & {\bf 1}$_{-4}$ & $2$ & $\sum_\al\Tr \Phi_\al^2$ \\
$T_4$ & {\bf 1}$_{-8}$ & $4$ & $\sum_\al\Tr \Phi_\al^4$ \\
      & {\bf 3}$_{-4}$ & $4$ & $\sum_\al\Tr \tilde Q_\al \Phi_\al^2 Q_\al$ \\
      & {\bf 5}$_0$ & $4$ & $\sum_\al\Tr ({\tilde Q}_\al Q_\al)^2$ \\
\hline
\end{tabular}\end{center}
\caption{Bulk spectrum of chiral primaries for $n=2$. The notation for $SU(2,2|2)$
multiplets is explained in the text.}
\end{table}

In Tables 2---4 we summarize the chiral primaries for ${\bf Z}_n$
orbifolds with $n>2$. The greater is $n$, the greater is the gap to the
next invariant state, therefore the less states we get in the
table.

\begin{table}\begin{center}
\begin{tabular}{|c|c|c|c|}
\hline 
$SU(2,2|2)$ & $SU(2)_R \times U(1)_R$ & $\Delta$ & Comments\\
\cline{1-4}
$G$ & {\bf 1}$_0$ & $2$ & \\
$V$ & {\bf 3}$_{0}$  & $2$ & \\
$T_3$ & {\bf 1}$_{-6}$ & $3$ & Modulus\\
      & {\bf 5}$_{0}$ & $4$ & \\
\hline
\end{tabular}\end{center}
\caption{Bulk spectrum of chiral primaries for $n=3$.}
\end{table}

\begin{table}\begin{center}
\begin{tabular}{|c|c|c|c|}
\hline 
$SU(2,2|2)$ & $SU(2)_R \times U(1)_R$ & $\Delta$ & Comments \\
\cline{1-4}
$G$ & {\bf 1}$_0$ & $2$ & \\
$V$ & {\bf 3}$_{0}$  & $2$ & \\
$T_4$ & {\bf 1}$_{-8}$ & $4$ & Modulus \\
      & {\bf 5}$_{0}$ & $4$ & \\
\hline 
\end{tabular}\end{center}
\caption{Bulk spectrum of chiral primaries for $n=4$.}
\end{table}

\begin{table}\begin{center}
\begin{tabular}{|c|c|c|}
\hline 
$SU(2,2|2)$ & $SU(2)_R \times U(1)_R$ & $\Delta$ \\
\cline{1-3}
$G$ & {\bf 1}$_0$ & $2$\\
$V$ & {\bf 3}$_{0}$ & $2$ \\
    & {\bf 5}$_{0}$ & $4$ \\
\hline 
\end{tabular}\end{center}
\caption{Bulk spectrum of chiral primaries for $n=6$.}
\end{table}

In the preceeding analysis we ignored states with nonzero $U(1)_L$ charge. There is
a good reason for this. Recall that the ${\bf Z}_{nk}$ subgroup of $U(1)_L$ is
part of the orbifold group; a ${\bf Z}_{nk}$-invariant state must have $U(1)_L$ charge
$L=0 $ mod $nk$. If $L\neq 0$, it must be of order $k$, therefore for large $k$
all such states come from $SU(4)$ representations with large dimension. Hence
for sufficiently large $k$ all states with $L\neq 0$ couple to irrelevant operators.
If one is only interested in relevant or marginal operators, one needs to consider
states with $L\neq 0$ only for a few low values of $k$. As an example, we  
consider the case $k=1$. The L-charged chiral primaries are shown in Table 5. For
$n=2$ we indicated the operators in the gauge theory the $AdS$ states couple to. As 
explained above, $U(1)_L$ is a symmetry of the gauge theory with respect to which
$\tilde Q$ and $Q$ have charges $1$ and $-1$, while the rest of the fields are neutral.
Note also that for $n=2,\ k=1$ the SCFT on the boundary has gauge group $Sp(N)\times Sp(N)$,
so the bifundamental representation is self-conjugate. The invariant
antisymmetric tensors of $Sp(N)\times Sp(N)$ which we denote by $J$ are used to
raise and lower gauge indices.

\begin{table}\begin{center}
\begin{tabular}{|c|c|c|c|c|}
\hline 
      & $U(1)_L$ & $SU(2)_R \times U(1)_R$ & $\Delta$ & SCFT operator\\
\cline{1-5}
$n=2$ & $2$      & {\bf 3}$_0$    & $2$ & $\Tr (QJ)^2$\\
      & $2$      & {\bf 3}$_{-4}$ & $4$ & $\Tr (QJ)^2\Phi^2$\\
      & $-2$     & {\bf 3}$_{-4}$ & $4$ & $\Tr \Phi^2 (J{\tilde Q})^2$\\
      & $2$      & {\bf 5}$_0$    & $4$ & $\Tr J\tilde Q (QJ)^3$\\
      & $4$      & {\bf 5}$_0$    & $4$ & $\Tr (QJ)^4$\\
\hline
$n=3$ & $3$      & {\bf 4}$_0$    & $3$ &  \\
\hline
$n=4$ & $4$      & {\bf 5}$_0$    & $4$ &  \\
\hline
\end{tabular}\end{center}
\caption{Chiral primaries with nonzero $U(1)_L$ charge for $k=1$.}
\end{table}

\subsection{${\bf Z}_{nk}$ twisted states}

The ${\bf Z}_{nk}$ orbifold has $nk-1$ twisted sectors labeled by $j=1,\ldots,nk-1$.
Each twisted sector contributes a $(2,0)$ tensor multiplet in $d=6$~\cite{DM}. 
It contains a 2-form with antiself-dual field strength, a complex scalar which is a 
singlet of $SU(2)_R$,
and three real scalars which transform as ${\bf 3}$ of $SU(2)_R$. Kaluza-Klein (KK) reduction 
of this tensor multiplet on a circle yields a tower of $SU(2,2|2)$ multiplets~\cite{g}.
Let us denote the KK momentum along the circle $\ell$.
The R-charge of the state is $R=2\ell$.
Recalling the definition of $\CP_n J_n$, we infer that the $\CP_n J_n$ projection
requires the states from the $s^{\rm th}$ sector to have R-charge $-2s$ mod $2n$.
The states originating from the complex scalars in $d=6$ are special because of the additional
monodromy: their R-charge is $-2s\pm 2$ mod $2n$. In either case, since $s$ runs from
$1$ to $nk-1$, it is convenient to write it as $s=j+np$, where $j=0,\ldots,n-1$,
while $p$ runs from $1$ to $k-1$ for $j=0$ and from $0$ to $k-1$ for all other $j$.
(We exclude $j=p=0$ because it is the bulk sector which we have already
analyzed).
This parametrization is convenient because it makes obvious that the R-charge mod $2n$
depends only on $j$, and not on $p$. Namely, states which do not come from
complex scalars have $R=-2j$ mod $2n$. It follows that the multiplicity of such a state is $k$ 
for $j\neq 0$ and $k-1$ for $j=0$. Similarly, the R-charge of the complex scalar states is 
$\pm 2-2j$ mod $2n$.

Let us start with the KK reduction of the 2-forms. In general one expects that a KK
reduction of a 2-form in $d=6$ yields a 2-form on $AdS^5$. Such states are always 
massive~\cite{gun}. 
The mass of a 2-form is related to the dimension of an operator on the boundary via
$m^2=(\Delta-2)^2$.
It is easy to see that $m^2=\ell^2$. 
The case of zero KK momentum is special: instead of a 2-form KK reduction yields a real
massless vector on $AdS^5$ (see Appendix). As explained above, only states from the 
$j=0$ sector are allowed to have zero KK momentum. Their multiplicity is $k-1$, 
for any $n$. Massless vectors couple to conserved currents of dimension $\Delta=3$.
Recalling that we also obtained a $U(1)_L$ gauge boson from the bulk sector, we conclude that the 
SCFT on the boundary has a total of $k$ $U(1)$ currents. This should be true independent of $n$.

Next consider the reduction of scalars. According to~\cite{g},
real $SU(2)$ triplets yield states on $AdS^5$ with mass $m^2=\ell^2-4$, while
complex singlets yield states with $m^2=\ell^2\pm 4\ell$. 
With these formulas at hand we can now find all twisted supergravity states which couple to 
relevant or marginal operators on the boundary. The results are summarized in Tables 6---8. 
Again we listed only chiral primaries. Besides the ultra-short vector multiplets which 
couple to $U(1)$ currents we also have tensor multiplets $T_{-\ell},\ \ell<0$. The
lowest component of $T_{-\ell}$ is a scalar with KK momentum $\ell$.

\begin{table}\begin{center}
\begin{tabular}{|c|c|c|c|c|}
\hline 
$SU(2,2|2)$ & Multiplicity & $SU(2)_R \times U(1)_R$ & $\Delta$ & SCFT operator\\
\cline{1-5}
$V$ & k-1 & {\bf 3}$_0$ & $2$ & $\tilde Q_\al Q_\al$ \\
$T_2$ & k & {\bf 1}$_{-4}$ & $2$ & $\Tr \Phi_\al^2$ \\
$T_3$ & k-1 & {\bf 1}$_{-6}$ & $3$ & $\Tr \Phi_\al^3$ \\
$T_4$ & k & {\bf 1}$_{-8}$ & $4$ & $\Tr \Phi_\al^4$ \\
\hline
\end{tabular}\end{center}
\caption{Chiral primaries from ${\bf Z}_{nk}$-twisted sectors for $n=2$.}
\end{table}

\begin{table}\begin{center}
\begin{tabular}{|c|c|c|c|c|}
\hline 
$SU(2,2|2)$ & Multiplicity & $SU(2)_R \times U(1)_R$ & $\Delta$ & Comments\\
\cline{1-5}
$V$   & k-1 & {\bf 3}$_0$ & $2$ & \\
$T_p$ & k & {\bf 1}$_{-2p},\ p=2,3$ & $p$ & Moduli\\
$T_4$ & k-1 & {\bf 1}$_{-8}$ & $4$ & Moduli \\
\hline
\end{tabular}\end{center}
\caption{Chiral primaries from ${\bf Z}_{nk}$-twisted sectors for $n=3$.}
\end{table}

\begin{table}\begin{center}
\begin{tabular}{|c|c|c|c|c|}
\hline 
$SU(2,2|2)$ & Multiplicity & $SU(2)_R \times U(1)_R$ & $\Delta$ & Comments\\
\cline{1-5}
$V$ & k-1 & {\bf 3}$_0$ & $2$ & \\
$T_p$ & k & {\bf 1}$_{-2p},\ p=2,3,4$ & $p$ & Moduli \\
\hline
\end{tabular}\end{center}
\caption{Chiral primaries from ${\bf Z}_{nk}$-twisted sectors for $n=4$ and $n=6$.}
\end{table}

\subsection{Comparison with Seiberg-Witten curves}
In this subsection we compare the supergravity states we found with the results of section 2.
First, let us compare the spectrum of $SU(2)_R$ singlet primaries with the Coulomb branch 
moduli. It was shown in section 2 that the Coulomb branch moduli have 
R-charge $4,6,8,\ldots,2Nn$. The number of moduli is determined by their R-charge:
\be\label{multR}
\# \ {\rm of\ moduli} =\left\{ \begin{array}{ll}
k+1, & R=0 \ {\rm mod}\  2n \\
k-1, & R=2 \ {\rm mod}\  2n \\
k & {\rm otherwise} \end{array} \right.
\ee
The dimensions of the moduli are determined by their R-charges as well, $\Delta=R/2$.
Since we limited the supegravity analysis to operators with $\Delta\leq 4$, we
can only compare the moduli with $R\leq 8$. A quick look at Tables 1---8 shows that
indeed all R-charges and multiplicities are in agreement with Eq.~(\ref{multR}). 

Second, according to section 2, the theories in questions have exactly marginal
deformations with zero R-charge corresponding to the locations of the M5$'$ branes
and the complex structure of ${\bf T}^2$. Their number is $k+1$ for $n=2$ and $k$ for $n>2$.
We suggest the following identification of these deformations on the supergravity
side. According to our tables, for $n=2$ we have $k+1$ $SU(2,2|2)$ multiplets
$T_2$, while for $n>2$ we have only $k$ such multiplets. The lowest components
of these multiplets couple to operators of dimension $2$ and have been identified above 
as the Coulomb branch moduli. Their highest components are complex scalars which transform as
{\bf 1}$_0$ with respect to $SU(2)_R\times U(1)_R$ and couple to operators of dimension $4$.
It is natural to interpret these complex scalars as the exactly marginal
deformations required by the Seiberg-Witten curve. For $n=2$ we can check this
identification by a gauge theory calculation. In this case the lowest components
are simply $\Tr \Phi_\al^2$, and the corresponding highest components are
$\Tr (F_\al^2+i F_\al\tilde F_\al)$. Deformations by $\Tr (F_\al^2+i F_\al{\tilde F}_\al)$
are equivalent to changing the gauge couplings and theta-angles. Such deformations
are indeed exactly marginal. Furthermore, in the brane construction of section 2
the complexified gauge couplings were encoded in the positions of the M5$'$ branes
on ${\bf T}^2$ and the complex structure of ${\bf T}^2$. 

Supergravity also predicts that the SCFT on the boundary has $k$
conserved $U(1)$ currents. One of them couples to
a $U(1)_L$ gauge boson from the bulk sector. The other $k-1$ gauge bosons
arise from the twisted sector 2-forms with zero KK momentum (see Appendix).
The presence of $k$ $U(1)$ currents is in agreement with the analysis of the
Seiberg-Witten curves in section 2. There we identified the mass deformations
as the lowest components of the background vector multiplets which couple 
to conserved currents. In particular, it is natural to identify the global mass
as the superpartner of the $U(1)_L$ current.

The conclusion is that the supergravity spectrum of $SU(2)_R$ singlet scalars  
matches precisely with the parameters of the Seiberg-Witten curve. The agreement
depends crucially on the fact that the F-theory orbifold action contains
a nongeometric part $J_n$. 

One interesting feature to note is how the
decoupling of $U(1)$'s in the gauge theory is manifested in the supergravity.
A signature of decoupling is that the 2-form operators $\Tr F_\al$ are absent and 
one gets global $U(1)$ currents instead. In the supergravity approach a trade-off 
between a 2-form and a gauge boson occurs because of a subtlety in the KK reduction
of the 2-form on $AdS^5\times S^1$: for zero KK momentum the reduction yields
a gauge boson on $AdS^5$, while for nonzero KK momentum it yields a massive 2-form.
Note also that all masses for the bifundamentals come from the fluxes of the 
B-fields through the shrunk 2-cycles of ${\bf C}^2/{\bf Z}_{nk}$. In the case of
the global mass parameter it is better to think about an ALF space
with a ${\bf Z}_{nk}$ singularity rather than about ${\bf C}^2/{\bf Z}_{nk}$. The 
global mass arises from the B-field flux through the noncompact 2-cycle
of the ALF.  

\subsection{$7$-brane excitations}

So far we considered states neutral with respect to the 7-brane gauge group $H$.
Let us now discuss the $H$-charged states. They
appear from sectors twisted with respect to $\CP_n J_n$ and are localized 
on 7-branes. For $n=2$ they are simply open string with both ends of the D7 branes.
For $n>2$ the twisted sector also includes multi-prong strings with ends on mutually
nonlocal 7-branes.

To compute the spectrum of 7-brane excitations it is convenient to think of the
orbifolding procedure in the following way. One starts with 7-brane states in $d=8$
with gauge group $G$. For $n=2,3,4,6$ $G=Spin(8)/{\bf Z}_2, E_6, E_7, E_8$, respectively.
All massless states in $d=8$ live in a vector multiplet in the adjoint of $G$ which
includes a vector and a complex scalar. The complex scalar is in the {\bf (1,1)}$_2$
of $SU(2)_L\times SU(2)_R\times U(1)_R$. 
Next one wraps the 7-branes on an orbifold ${\bf C^2}/{\bf Z}_{nk}$. 
As explained in section 3, the orbifold group acts nontrivially on the 7-brane gauge 
bundle and breaks $G$ to a subgroup $H$. The supergravity spectrum is obtained by
reducing the $d=8$ spectrum to $d=5$, decomposing the adjoint of $G$ into
representations of $H$, and then projecting on the states invariant
with respect to the ${\bf Z}_{nk}$ action. 
The resulting $d=5$ spectrum depends on the manner in which $G$
is broken down to $H$. Since we know $H$ only for $n=2$, we first investigate this
case.

For $n=2$ the orbifold action on $G$ is given by Eq.~(\ref{monodr}). 
The unbroken gauge group is
$H=(Spin(4)\times Spin(4))/{\bf Z}_2$. The adjoint of $G=Spin(8)/{\bf Z}_2$
decomposes into an adjoint and a pair of {\bf (4,4)} representations of $H$. The adjoint
of $H$ commutes with $\CM$, while the {\bf (4,4)} representations
anticommute with it. As explained in~\cite{afm}, the
reduction of the $d=8$ vector multiplet to $d=5$ produces a tower of vector multiplets
whose lowest components are real scalars in {\bf (p,p+2)}$_0$ of $SU(2)_L\times SU(2)_R
\times U(1)_R$, and $p=1,2,3,\ldots.$ The scalars are chiral primary
states, so their masses are given by $m^2=(p+1)(p-3)$. For $p=1$ the vector
multiplet contains a massless vector, while for $p>1$ it contains a massive vector.
The requirement that the scalar couple to a relevant or marginal
operator on the boundary restricts $p$ to be $1,2,$ or $3$.

To perform the ${\bf Z}_{2k}$ projection we recall that the geometric part of
${\bf Z}_{2k}$ is embedded into
$U(1)_L\subset SU(2)_L$. First let us look at states with zero $U(1)_L$ charge $L$.
Such states are present only for odd $p$.
For these states only the nongeometric part of ${\bf Z}_{2k}$ is nontrivial.
This means that the {\bf (4,4)} representation of $H$ is projected out and only the
adjoint of $H$ remains. 

Turning now to states with $L\neq 0$, it is
easy to see that unless $k=1$ or $2$ they only couple to irrelevant operators. Indeed, 
invariance with respect to the ${\bf Z}_k$ subgroup of ${\bf Z}_{2k}$
requires that $L$ be a multiple of $k$. But
the states which couple to relevant or marginal operators (i.e. states with $p=1,2,3$) have
$|L|\leq 2$. Thus we only need to consider $k=1$ and $k=2$.
 
For $k=1$ we need to perform a ${\bf Z}_2$ projection. The generator of ${\bf Z}_2$ 
is a product of a rotation by $\pi$ in $U(1)_L$ and a conjugation by $\CM$. 
The $p=2$ scalar transforms
as {\bf (2,4)}$_0$ of $SU(2)_L\times SU(2)_R \times U(1)_R$, so 
it is odd with respect to the $\pi$ rotation. Hence its ${\bf Z}_2$-invariant part is
a complex scalar in the {\bf (4,4)} of $H$ which transforms as {\bf 4}$_{0,+2}$ of 
$SU(2)_R\times U(1)_R\times U(1)_L$. The $p=3$ scalar transforms as 
{\bf (3,5)}$_0$ of $SU(2)_L\times SU(2)_R \times U(1)_R$, and its ${\bf Z}_2$-invariant part
includes a complex scalar
in the adjoint of $H$ and in {\bf 5}$_{0,+2}$ of $SU(2)_R\times U(1)_R\times U(1)_L$.
(It also includes a {\bf 5}$_0$ scalar with $L=0$ which we have already discussed.)

For $k=2$ we need to perform a ${\bf Z}_4$ projection. The generator of ${\bf Z}_4$
is a product of a rotation by $\pi/2$ in $U(1)_L$ and a conjugation by $\CM$. 
The $p=2$ scalar is projected out completely, while the $p=3$ scalar yields a
complex scalar in the {\bf (4,4)} of $H$ and in 
{\bf 5}$_{0,+2}$ of $SU(2)_R\times U(1)_R\times U(1)_L$.

The results of this analysis are summarized in Table 9.
We matched all the states we found with gauge theory operators. The gauge group for $n=2$
contains two symplectic factors $Sp(N)_1$ and $Sp(N)_2$ with two fundamentals for
each factor. In Table 9 $q_\sigma, \tilde q_\sigma, \sigma=1,2,$ denote the 
scalars in the fundamental hypermultiplet of $Sp(N)_\sigma$.

\begin{table}\begin{center}
\begin{tabular}{|c|c|c|c|c|c|l|}
\hline
H       & $U(1)_L$ & $SU(2)_R\times U(1)_R$ & $\Delta$ & SCFT operator & Comments\\
\cline{1-6}
Adjoint & $0$    & {\bf 3}$_0$            & $2$      & $\tilde q_\sigma q_\sigma$ & \\
Adjoint & $0$    & {\bf 5}$_0$            & $4$      & $\tilde q_\sigma Q {\tilde Q} q_\sigma$ &\\
{\bf (4,4)} & $1$  & {\bf 4}$_0$ & $3$ & $\tilde q_2 JQJ q_1$ & $k=1$ \\
Adjoint & $2$ & {\bf 5}$_0$ & $4$ & $\tilde q_\sigma (QJ)^2 q_\sigma$ & $k=1$\\
{\bf (4,4)} & $2$ & {\bf 5}$_0$ & $4$ & $\tilde q_2 Q_2 Q_1 J q_1$ & $k=2$\\
\hline
\end{tabular}\end{center}
\caption{States charged with respect to the 7-brane gauge group $H$ for $n=2$. Some
of the states are present only for special values of $k$; this is indicated in the
last column.}
\end{table}

Finally let us discuss the 7-brane states for $n=3,4,6$. The analysis of states
with zero $U(1)_L$ charge proceeds in exactly the same way as for $n=2$ and
produces the same spectrum of states. All these states live in the adjoint of $H$.
The only difference from the case $n=2$ is that now we do not know the monodromy $\CM$,
and so do not know $H$. Again it is easy to see that for $k>2$ the states with 
$L\neq 0$ couple only to irrelevant operators. To analyze the spectrum
of states with $L\neq 0$ for $k=1,2$ one needs to know $\CM$.

\section{Conclusions and outlook}
In this paper we constructed and studied a class of 
$\CN=2$ superconformal field theories in four dimensions. These theories
are labeled by two integers $n$ and $k$. $n$ takes values $2,3,4,6$, and $k$
is an arbitrary positive integer. For $n=2$ the model in question is a finite
$\CN=2$ gauge theory with gauge group $Sp(N)\times SU(2N)^{k-1}\times Sp(N)$.
For $n>2$ the models are new $\CN=2$ field theories which do not admit a 
Lagrangian description. Using M and F-theory methods we learned quite a lot
about these theories: we determined the Seiberg-Witten curve and found the
spectrum of operators in short representations of the superconformal group.
The M-theory approach allows to study the theories for finite $N$ but is
limited to superconformal families containing the Coulomb branch moduli.
To learn about other families we used F-theory and the AdS/CFT correspondence. This
approach  is effective for large $N$. The F-theory construction turns out
to be quite intricate and involves frozen ${\bf C}^2/{\bf Z}_n$ singularities.
It is satisfying to see that whenever both methods apply they give identical
results. In particular, we showed that the decoupling of $U(1)$'s in the gauge
theory on the boundary of $AdS^5$ is due to a subtlety in the KK reduction of
a 2-form on $AdS^5\times S^1$. 

An interesting direction to pursue is to study in more detail F-theory backgrounds
with frozen orbifold singularities. Only in the case $n=2$, where the background can
be understood as an orientifold background in IIB, do we have a complete control
over the twisted sectors of the orbifold. For $n>2$ the twisted sectors must include
multi-prong strings. It would be quite interesting to study the structure of the
twisted sectors in detail and in particular determine the gauge group living at
the singularity. It is likely that this can be done along the lines of~\cite{sing}.

One could extend our results by considering 7-branes wrapping
orbifold singularities other than $A_n$. For example, one could consider D7-branes wrapping
a $D_n$-type singularity. In M-theory this corresponds to adding 
``NS orientifolds''~\cite{Dq} parallel to NS5 branes. The Seiberg-Witten solution for
the probe theory will again be encoded in a Hitchin system on an orbifolded torus, 
the main difference being that the gauge group of Hitchin equations will be $SO(2N)$ rather
than $U(2N)$.

\renewcommand{\thesection}{Appendix:}
\section{Kaluza-Klein reduction of 2-form on $AdS^5\times S^1$}

Consider a 2-form $B$ on $AdS^5\times S^1$ whose field strength
$G=dB$ satisfies $G=-*G$. We are going to show that the Kaluza-Klein
reduction of $B$ produces either a massless vector on $AdS^5$ satisfying
Maxwell equations, or a massive 2-form satisfying equations of the anti-selfdual
type, depending on whether the momentum along $S^1$ is zero or not.

The Kaluza-Klein ansatz is
\be
B=(a\wedge dt+b)e^{it\ell},
\ee
where $a$ is a 1-form on $AdS^5$, $b$ is a 2-form on $AdS^5$, and $\ell$ is
the integer-valued KK momentum. Then $G$ and $*G$ are given by
\be
G=(da\wedge dt+db+i\ell b\wedge dt)e^{it\ell},\qquad 
*G=(*da+*db\wedge dt+i\ell *b)e^{it\ell}.
\ee
The equation $G=-*G$ implies an equation of motion for $a$ and $b$:
\be
da+i\ell b=-*db.
\ee
The field $B$ in $d=6$ has a gauge invariance $B\ra B+d\Lambda$, where $\Lambda$ is a 1-form.
Kaluza-Klein ansatz is invariant with respect to a subset of these
gauge transformations, namely those with $\Lambda$ of the form
\be
\Lambda=(\sigma dt+ \lambda)e^{it\ell}.
\ee
Here $\sigma$ and $\lambda$ are 0- and 1-form on $AdS^5$, respectively.
The induced transformations on $a$ and $b$ are
\be
a\ra a+d\sigma+i\ell\lambda,\qquad b\ra b+d\lambda.
\ee

When $\ell=0$ the gauge invariance for $a$ is the standard gauge invariance for
the massless vector field, $a\ra a+d\sigma$. The equation of motion takes
the form $da=-*db$. This is equivalent to Maxwell equations $d*da=0$.
Then $b$ is not an independent field: it is determined by $a$ up to a gauge
transformation.

When $\ell\neq 0$ gauge freedom can be used to set $a=0$. Then
$b$ does not have any residual gauge invariance. Its equation of motion
becomes $i\ell b=-*db$. This is an equation of the anti-selfdual type
describing a massive 2-form on $AdS^5$~\cite{krn,gun}.
  

\section*{Acknowledgements}

The authors wish to thank O.~Aharony, M.~Berkooz, and E.~Witten for helpful discussions.
The work of S.G. was supported in part by NSF grant PHY-9802484, RFBR grant No
98-02-16575, and Russian President's grant No 96-15-96939.
The work of A.K. was supported in part by DOE grant DE-FG02-90-ER40542.


\begin{thebibliography}{99}

\bibitem{SWAP} P.C. Argyres, M.R. Plesser, N. Seiberg, and E. Witten,
``New N=2 Superconformal Field Theories in Four Dimensions,''
Nucl. Phys. {\bf B461}, 71-84 (1996).

\bibitem{maldacena} J. Maldacena, ``The Large N Limit of Superconformal 
Field Theories and Supergravity,'' hep-th/9711200.

\bibitem{AdS} S.S. Gubser, I.R. Klebanov, and A.M. Polyakov,
``Gauge Theory Correlators From Noncritical String Theory,'' hep-th/9802109;
E. Witten, ``Anti-de Sitter Space and Holography,'' hep-th/9802150.

\bibitem{Angel} A.M.~Uranga, ``Towards mass deformed ${\cal N}=4$ $SO(n)$
and $Sp(k)$ theories from brane configurations,'' Nucl. Phys. {\bf B526}, 241-277 (1998),
hep-th/9803054.

\bibitem{Witten} E. Witten, ``Solutions Of Four-Dimensional Field Theories Via M
Theory,'' Nucl. Phys. {\bf B500}, 3-42 (1997).

\bibitem{DM} M.~Douglas and G.~Moore, ``D-Branes, Quivers, and ALE Instantons,''
hep-th/9603167.

\bibitem{DasMu} K. Dasgupta and S. Mukhi, ``F-theory at Constant Coupling,''
Phys. Lett. {\bf B385}, 125-131 (1996), hep-th/9606044.

\bibitem{P} J. Polchinski, ``Tensors from K3 Orientifolds,'' hep-th/9606165.

\bibitem{K} A. Kapustin, ``Solution of N=2 Gauge Theories via
Compactification to Three Dimensions,'' hep-th/9804069.

\bibitem{SW3} N.~Seiberg and E.~Witten, ``Gauge Dynamics and
Compactification to Three Dimensions,'' hep-th/9607163.

\bibitem{IB} K. Intriligator, ``RG Fixed Points in Six Dimensions via Branes
at Orbifold Singularities,'' Nucl. Phys. {\bf B496}, 177-190 (1997), hep-th/9702038;
J.D. Blum and K. Intriligator, ``Consistency Conditions for Branes at
Orbifold Singularities,'' Nucl. Phys. {\bf B506}, 223-235 (1997),
hep-th/9705030.

\bibitem{VVW} C. Vafa, ``Modular Invariance and Discrete Torsion on Orbifolds,''
Nucl. Phys. {\bf B273}, 592 (1986);
C. Vafa and E. Witten, ``On Orbifolds with Discrete Torsion,'' J. Geom.
Phys. {\bf 15}, 189-214 (1995), hep-th/9409188.

\bibitem{HZ} A.~Hanany and A.~Zaffaroni, ``Branes and Six-Dimensional Supersymmetric Theories,''
hep-th/9712145.

\bibitem{KS} A.~Kapustin and S.~Sethi, ``The Higgs Branch Of Impurity
Theories,'' hep-th/9804027.

\bibitem{GLY} B.R.~Greene, C.I.~Lazaroiu, and P.~Yi, ``D Particles on $T^4/Z_N$
Orbifolds and Their Resolutions,'' hep-th/9807040.

\bibitem{DW} R. Donagi and E. Witten, ``Supersymmetric Yang-Mills Theory And
Integrable Systems,'' Nucl. Phys. {\bf B460}, 299-334 (1996).

\bibitem{sen} A.~ Sen, ``F-theory and Orientifolds'',
Nucl.Phys. {\bf B475} (1996) 562-578.

\bibitem{probe} M.R.~Douglas, D.A.~Lowe, and J.H.~Schwarz, ``Probing F-theory with
Multiple Branes,'' Phys. Lett. {\bf B394}, 297-301 (1997), hep-th/9612062.

\bibitem{six} M. Berkooz et al., ``Anomalies, Dualities, and Topology of
$D=6\ N=1$ Superstring Vacua,'' Nucl. Phys. {\bf B475}, 115-148 (1996).

\bibitem{SS} A.~Sen and S.~Sethi, ``The Mirror Transform of Type I Vacua in Six Dimensions,''
Nucl. Phys. {\bf B499}, 45-54 (1997), hep-th/9703157.

\bibitem{GP} E.G. Gimon and J. Polchinski, ``Consistency Conditions for
Orientifolds and D-Manifolds,'' Phys. Rev. {\bf D54}, 1667-1676 (1996), hep-th/9601038.

\bibitem{GZw} M.R.~Gaberdiel and B.~Zwiebach, ``Exceptional Groups from Open Strings,'' Nucl. Phys.
{\bf B518}, 151-172 (1998), hep-th/9709013.

\bibitem{toz} Y.~Oz, J.~Terning, ``Orbifolds of $AdS_5 \times S^5$
and 4d Conformal Field Theories,'' hep-th/9803167.

\bibitem{fs} A.~Fayyazuddin, M.~Spalinski, ``Large N Superconformal Gauge
Theories and Supergravity Orientifolds,'' hep-th/9805096.

\bibitem{afm} O.~Aharony, A.~Fayyazuddin, J.~Maldacena, ``The Large N
Limit of ${\cal N}=2,1$ Field Theories from Threebranes in
F-theory,'' hep-th/9806159.

\bibitem{krn} H.J.~Kim, L.J.~Romans, P.~van~Nieuwenhuizen, ``Mass Spectrum of
Chiral Ten-Dimensional N=2 Supergravity on $S^5$,'' Phys. Rev. {\bf D32},
389-399 (1985).

\bibitem{gun} M.~Gunaydin, L.J.~Romans, N.P.~Warner, ``Compact and
Non-Compact Gauged Supergravity Theories in Five Dimensions,'' 
Nucl. Phys. {\bf B272}, 598 (1986).

\bibitem{g} S.~Gukov, ``Comments on N=2 AdS Orbifolds,'' hep-th/9806180.

\bibitem{sing} M.~Bershadsky, K.~Intriligator, S.~Kachru, D.R.~Morrison,
V.~Sadov, and C.~Vafa, "Geometric Singularities and Enhanced Gauge Symmetries",
Nucl. Phys. {\bf B481}, 215 (1996).

\bibitem{Dq} A.~Kapustin, ``$D_n$ Quivers From Branes,'' hep-th/9806238.

\end{thebibliography}
\end{document}